# Burnback Analysis of Solid Propellant Rocket Motors

Juan M. Tizón

E-mail: jm.tizon@upm.es

ORCID-ID: https://orcid.org/0000-0002-8687-6657

Departamento de Mecánica de Fluidos y Propulsión Aeroespacial, Escuela Técnica Superior de Ingeniería Aeronáutica y del Espacio (ETSIAE), Universidad Politécnica de Madrid (UPM), Pza. del Cardenal Cisneros 3, 28040 Madrid, Spain

# 1. Abstract

Burnback analysis is a geometric exercise, whose correct solution leads to obtaining the thrust curve of solid propellant rockets. Traditionally, Piobert's statement, which introduces a certain amount of intuition, is used as an argument to construct analytical and numerical algorithms, although it is also common to use numerical integration of differential equations, whose solution is free of ambiguities. This paper presents a detailed study of the process experienced by the combustion surface that allows enunciating the properties of the kinematics of the surface without the need to appeal to heuristic considerations. To the author's knowledge, although simple and usual in other disciplines, this kind of analysis has not been presented previously in the field of the combustion process of a solid propellant. A formal development of the theory allows us to identify the Eikonal equation as representative of the physical process and the one that is necessary to solve to obtain a true problem description. Next, the methods used throughout the technological development of solid propellant rockets are reviewed, from their beginnings, in which only analytical procedures and, at most, their automation were possible by means of the first calculators, to modern methods, which obtain solutions to complex problems, based on the numerical solution of PDE. Other methods are also reviewed, which are developed around some of the properties presented by the solution, that is, methods of heuristic or phenomenological foundation. As a result of the review, it becomes clear that the solution of the Eikonal equation for burnback analysis is undertaken in the early 2000's, clarifying the problem. However, all subsequent developments, systematically, employ techniques based on the Level Set Method developed in the late 1990s. But LSM is applied to much more general and complex problems, and its use adds nothing new to the problem solution. Finally, several examples of the capabilities of the most relevant methods are provided, from the point of view of both efficiency and precision, presenting results in situations of interest, in the field of propulsion by solid-propellant rockets.

# 2. Introduction

Solid-propellant rocket motors are the simplest high-performance propulsion system ever devised. It consists of a structural vessel filled with a mixture of energetic solid components, which react chemically at a high rate. This reaction produces gases at high temperature and pressure, which are expelled at high speed through a nozzle, producing the consequent reaction force, that is, thrust. When the solid propellant ignites and a combustion front is formed on its surface, it is gradually consumed layer by layer. The combustion geometry determines the propulsive response of the system, as it directly controls the mass released. By properly sizing the initially exposed area and anticipating what its variation will be, the thrust variation capacity is anticipated in the geometric design of the propellant (*throttling by design*).

From the economic point of view, solid propellant rocket engines are very effective propulsion systems due to the simplicity of their configuration and the ease and safety in the tasks of handling, transport, and use. From the propulsive point of view, the specific impulse they provide is modest, but in many of the space and terrestrial applications this weakness is compensated by simplicity in design and manufacturing economy. In addition, the solid propellant rocket motor has a very interesting impulse-density value that makes them the ideal system in applications where the volume is limited. To ensure the effective use of these systems and the fulfillment of the demanding requirements of the missions in which they are used, design and simulation tools with high degree of fidelity are necessary. In this sense, prediction of the thrust curve of the engine is essential. And, for this task, one must have versatile, fast, reliable, and accurate tools for analyzing the evolution of the combustion surface.

The calculation of the burning surface area as a function of time is an essential step in the analysis and design activities of solid propellant rocket engines. It is relatively easy to establish a heuristic procedure, based on a set of simple rules, that determine the evolution of the combustion surface with time for simple geometries, but only by a rigorous procedure can realistic and complex problems be addressed: for any initial geometry, or when the combustion rate is not constant.

Towards the third decade of the nineteenth century the French general of artillery Guillaume Piobert (1793-1871), military engineer and scientist, enunciated a hypothesis about what was the process that followed the combustion of the substances used in the impulsion of projectiles: *The combustion of the inner parts of the gunpowder grains takes place only when the layers that cover them are consumed; the speed with which the fire spreads from one cut to another, in the compound, has great influence on the effects of the explosion* (in his own words: "Rapidité de combustion. - La combustion des parties intérieures des grains de poudre n'a lieu que lorsque les couches qui les recouvrent sont consumées; la rapidité avec laquelle le feu se propage de tranche en tranche, dans la composition, a la plus grande influence sur les effets de l'explosion", this quote is from the publication *Mémoires sur les pouvoirs de guerre des différents procédés de fabrication: avec résumés des épreuves comparatives faites sur ces poudres à Esquerdes en 1831 et 1832 et à Metz en 1836 et 1837* , printer-bookseller Bachelier, 1844, Paris). That is, the propellant undergoes a local process, over the surface, and can be described by a combustion front that consumes it by layers, sequentially. If the rate of combustion is uniform, the layers have uniform thickness and the description of the evolution of the surface is reduced to a geometric calculation, in which the time variable is proportional to the depth advanced by the front.

In Figure 1, the photo corresponding to the geometry of a propellant in intermediate combustion times is presented. To obtain the images it is necessary to quench the motor (a procedure can be the sudden opening of the chamber, which causes a marked decrease in pressure that has as a consequence that the chemical reaction freezes, stopping the process of consumption of the solid). The initial geometry is an eight-pointed star. In the central photo the tips have been consumed, and the advance of the combustion front has also continued in the valleys. Finally, in the last photograph, taken close to the final moment, the combustion front is about to reach the engine casing, even though this will happen earlier at some points than at others. All of these features are a direct consequence of the initial geometry. Many of the aspects discussed in the preceding description have a marked influence on the performance of the system. The geometry with edges, the complete consumption of geometric entities (the tips) or the uneven consumption of the propellant that does not reach the casing simultaneously are indicators that determine the efficiency of the process.

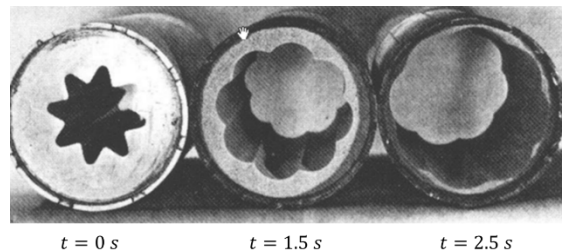

Figure 1: Situation of the combustion surface in three instants, the initial one, an intermediate state, and shortly before finishing the combustion process.

In practice, with uniform surface recession rate (idealized situation in which the pressure of the chamber must be uniform and the erosive combustion effects non-existent) the calculation of the evolution of the combustion surface involves its displacement perpendicular to itself. That is, each point on the surface is projected to a point on the new surface along the line perpendicular to the original surface. The normal distance traveled by the combustion front at each point will be called *forward coordinate* (symbol $y$). In the situation of constant burning rate, the value of the forward coordinate is proportional to the burning time. In this text, the term *pseudotime* is used (symbol $\tau$) when calculations are made with recession velocity equal to unity in the system of units in which the geometry of the propellant has been stored.

This has led to burnback analysis being approached on many occasions through analytical procedures with heuristic foundations, as in the well-known SPP© program [1], in which the initial surface is formed by extracting simple geometric elements from the volume of the chamber, such as parallelepipeds, spheres or cones whose combination and evolution reproduces the movement of a complex surface. However, the complexity of some combustion surfaces and the possibility of the process not taking place with constant recession velocity, make it advisable to establish a well-founded general analysis that allows the problem to be addressed in any situation.

Discrete methods should be used to assess the evolution of combustion surfaces in a general and automatic manner. Although analytical methods can be very quick and immediate, their application to complex geometries becomes complicated and laborious, or even unapproachable. Discrete methods offer the possibility of representing arbitrary combustion surfaces and delivering results automatically and repetitively. Usually, in the relevant literature, emphasis is placed on whether the methods solve the problem quickly or not, that is, whether they are computationally efficient. This interest is motivated because some methods employ search algorithms, which can slow them down if special precautions are not taken, and others involve the numerical integration of differential equations. Today, this aspect is of less relevance, because the power of computers has suffered a spectacular increase in recent years and because the impact of the method used in burnback analysis is small, on a global calculation of the design tasks. For application in the current context, the algorithms used to calculate the combustion surface at different times must be flexible, reliable, robust, and accurate. Flexible in the sense of allowing discretization of any surface and treatment of variable recession velocity. Effective and robust when calculating solutions in which interference effects may appear, such as caustics and rarefactions. And finally, accurate, which in principle could be regarded as a consequence of the previous but is also achieved by using adequate algorithms and well-founded mesh studies. In problems closely coupled with the resolution of the internal aerodynamics of the engine, the calculation time of the combustion area can be a non-negligible fraction of the total time, but an inert scalar in a domain of similar size should not exceed the fraction corresponding to the advection calculation. In addition, the calculation of the combustion surface must not contain many mesh points, when compared with those required in the detailed solution of a fluid field.

## **3.** Combustion front kinematics

Mathematically, the problem is to determine the function $S(x, y, z) - t = 0$ that describes the combustion surface at each point in time, in the domain initially occupied by the propellant, $x, y, z \in D$, where $t \geq 0$ is the time elapsed since ignition. It can also be said that the expression allows us to calculate the time ($t$) it takes for the combustion front to reach the point $(x, y, z)$ at which, naturally, $S(x, y, z) = 0$ defines the initial surface. For the correct approach to the problem, it is necessary to provide sufficient information about the value of the burning rate at each point, and that means knowing the recession velocity at all points of the volume initially occupied by the propellant, although its calculation is a consequence of the geometry at each instant.

Piobert's statement establishes that the combustion surface moves in the normal direction and suggests that each point on the surface moves perpendicular to the surface itself, but what happens is that the points disappear. The intuition of the scientist was correct, but it is worth developing a procedure that can be followed with confidence in any situation. To do this, imagine that we can refer to each point of the combustion surface $S(x, y, z) = t$ by means of a position vector, $\vec{r}_s(u, v, t)$ where $u$ and $v$ are two parameters, without specific physical dimensions, whose variation defines the surface. Now, it is assumed that both parameters define the surface in the region of interest with values of order unity, $u \sim v \sim 1$, although sometimes it may be convenient to parameterize the surface using the arc lengths, which will be expressly indicated. All points on the surface are subjected to the combustion process simultaneously and the geometry obtained is a consequence of this on the region occupied by the propellant (for example, $S \geq t \cap D$). To correctly analyze the problem, we will use the Huygens–Fresnel principle, which states that each point of a wavefront acts as a source point of a spherical wavefront, and that the interaction of all of them forms the propagation of the original front. Consider that the combustion process will affect only one point, $P$, at which the combustion process begins, as shown in Figure 2, and that the burning rate is constant and of value $\dot{r}_p$. After a time $\delta t$ the material consumed will be the one inside the intersection between the propellant and the sphere of center $P$ and radius $\dot{r}_p \delta t$. If it is now considered that all points on the surface of the propellant participate in the combustion process, each of them will be the center of a sphere that will have consumed the propellant inside. Over time the propellant contained inside all spheres will have been consumed and the combustion surface will be the envelope of the family of spheres internal to the propellant. This is a generalized algorithm for the determination of the new position of the combustion surface that can be applied whatever the shape of the combustion surface and that helps to solve any complicated configuration.

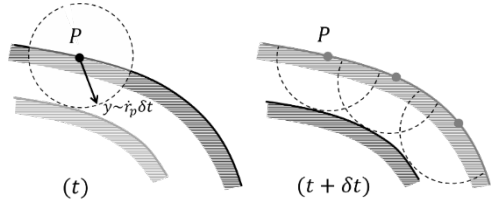

Figure 2: Diagram of the application of the Huygens–Fresnel principle to the determination of the motion of the combustion surface.

The family of spheres that have their center at a point on the surface $S$ and radio $\dot{r}_p \delta t$ is:

$$(\vec{r} - \vec{r}_s) \cdot (\vec{r} - \vec{r}_s) = (\dot{r}_p \delta t)^2 \qquad (1)$$

The envelope of the family is obtained by canceling out the derivative of the equation of the surface with respect to the parameters $u$ and $v$, and solving the generated system together with the equation of the family itself (1). If, in general, it is assumed that the burning rate depends on the position, differentiating yields

$$\frac{\partial \vec{r}_s}{\partial u} \cdot (\vec{r} - \vec{r}_s) = -(\dot{r}_p \delta t)^2 \frac{1}{\dot{r}_p} \frac{\partial \dot{r}_p}{\partial u} \qquad (2)$$

$$\frac{\partial \vec{r}_s}{\partial v} \cdot (\vec{r} - \vec{r}_s) = -(\dot{r}_p \delta t)^2 \frac{1}{\dot{r}_p} \frac{\partial \dot{r}_p}{\partial v} \qquad (3)$$

Replacing the parameters $u$ and $v$, equations (1), (2) and (3) provide the expression of the new surface. Note that the evolution of the combustion surface must be smooth, at least, in this development. As the combustion surface in time $t + \delta t$ is arbitrarily close to the original, using $\delta \vec{r} = \vec{r} - \vec{r}_s$, it is obtained that $|\delta \vec{r}| \sim \dot{r}_p \delta t$, according to equation (1), which is small compared to the characteristic size of the combustion surface $L \gg |\delta \vec{r}|$. Moreover, the left-hand side of equations (2) and (3) is of the order of $L|\delta \vec{r}|$, while the right-hand side is of the order of $|\delta \vec{r}|^2$, and since $|\delta \vec{r}|^2 \ll L|\delta \vec{r}|$, equations (2) y (3) must be replaced by

$$\partial \vec{r}_s/\partial u \cdot (\vec{r} - \vec{r}_s) = 0 \tag{4}$$

$$\partial \vec{r}_s/\partial v \cdot (\vec{r} - \vec{r}_s) = 0 \tag{5}$$

Consequently, the equations to be solved are (1), (4) and (5). Vectors $\partial \vec{r}_s/\partial u$ and $\partial \vec{r}_s/\partial v$ are tangent to the surface $S$ and it is concluded that, both $\partial \vec{r}_s/\partial u \cdot (\vec{r} - \vec{r}_s) = 0$ and $\partial \vec{r}_s/\partial v \cdot (\vec{r} - \vec{r}_s) = 0$, are the equations of planes perpendicular to the tangent vectors at the point $P$.

The above result cannot be applied on a combustion surface where the normal direction is not defined, but the algorithm of the sphere family does not require the surfaces to be smooth and is very useful when analyzing the evolution of the combustion surface in non-regular situations, with geometric elements such as cusps or corners. Also, it is possible to easily analyze complex situations, for example, conductive cables embedded in the propellant or bipropellant situations, with different burning rates.

The direction of advance of the surface is perpendicular to the surface $S$ and therefore parallel to the gradient vector, $\nabla S$. The modulus of the vector is related to the speed of advance of the front since, by the expression chosen at the beginning of this section, $S(x, y, z) - t = 0$, and in this way $\delta S = \delta t$ or, what is the same,

$$|\nabla S| = 1/\dot{r}_p \tag{6}$$

Which is known as the *Eikonal equation* (word that, in Greek, means "image"). This equation is basic in Geometric Optics because it allows the calculation of the trajectories of light rays, perpendicular to the surfaces of the same optical path and, therefore, the calculation of the trajectories that reverse a minimum time (Fermat's principle). In this case, the inverse of the burning rate plays the role of the refractive index (ratio between the light speed in vacuum and that of the medium). This equation is used not only in geometric optic applications, but also in other wave propagation problems, such as electromagnetism or seismology. The solutions of the equation can exhibit geometric singularities called *caustic* ("causticus" in Latin means "burnt") or the well-known mirage phenomenon.

The vector $d\vec{r}$ has the direction of $\nabla S$ and the modulus is the variation of the forward normal coordinate, $dy = \dot{r}_p dt$, with which equation (6) can be written as

$$\nabla S = \frac{1}{\dot{r}_p} \frac{d\vec{r}}{dy} \tag{7}$$

Differentiating with respect to $y$,

$$\frac{d}{dy}[\nabla S] = \frac{d}{dy}\left[\frac{1}{\dot{r}_p}\frac{d\vec{r}}{dy}\right] \tag{8}$$

The left-hand side can be transformed by the chain rule as follows,

$$\frac{d}{dy}[\nabla S] = \frac{d\vec{r}}{dy} \cdot \nabla[\nabla S] = \dot{r}_p \nabla S \cdot \nabla[\nabla S] = \frac{1}{2}\dot{r}_p \nabla[\nabla S \cdot \nabla S] = \nabla\left(\frac{1}{\dot{r}_p}\right) \tag{9}$$

And equation (8) becomes

$$\frac{d}{dy}\left[\frac{1}{\dot{r}_p}\frac{d\vec{r}}{dy}\right] = \nabla\left(\frac{1}{\dot{r}_p}\right) \tag{10}$$

The equation with which the trajectory of the surface points can be calculated. Developing the derivatives yields

$$\frac{1}{\dot{r}_p}\frac{d\dot{r}_p}{dy}\frac{d\vec{r}}{dy} - \frac{d^2\vec{r}}{dy^2} = \frac{\nabla\dot{r}_p}{\dot{r}_p} \tag{11}$$

From which certain interesting properties can be derived. The first one is that, if the burning rate is constant, the surface points move along straight lines since the solution of $d^2\vec{r}/dy^2 = 0$, is

$$\vec{r} = \vec{r}_0 + y(\nabla S/|\nabla S|) \tag{12}$$

Where $\vec{r}_0$ is the starting position and it has been used that $\dot{r}_p|\nabla S| = 1$. On the other hand, by construction, $d\vec{r}/dy$ is a vector in the direction of the normal to the surface, whereas $d^2\vec{r}/dy^2$ is perpendicular to it, so that the recession rate gradient can be broken down into a normal component $\nabla_\perp \dot{r}_p = d\dot{r}_p/dy$ and a parallel component $\nabla_\parallel \dot{r}_p$. Equation (11) can therefore be projected in the directions perpendicular and parallel to the surface. In the direction perpendicular to the surface the result is trivial (equation (12)) while in the parallel direction

$$\frac{d^2\vec{r}}{dy^2} = -\frac{\nabla_\parallel \dot{r}_p}{\dot{r}_p} \tag{13}$$

Which expresses that the trajectories only turn if there is a non-zero parallel gradient of the recession rate. When the recession rate is constant the combustion surface can be reconstructed by simple translations. For this reason, numerous heuristic algorithms have been developed over time to solve this problem.

Some general results, related to geometric optics, of interest for the performances of rocket engines have been reviewed. But the relevant thing is to calculate the combustion area at each moment, because it allows us to determine the thrust curve. To have a means of assessing the area of combustion, the surface must be parameterized with $\vec{r}_s(u, v, t)$ (note that the subscript will be ignored hereafter), assuming that the values of $u$ y $v$ identify a point on the surface and, as long as the value of the parameters is maintained, the point follows the trajectory described by (11). In other words, parameterization complies with

$$\frac{\partial \vec{r}}{\partial t} = \dot{r}_p \vec{n} \tag{14}$$

Where the normal to the surface $\vec{n}$ is calculated as usual,

$$\vec{n} = \frac{\vec{r}_u \times \vec{r}_v}{|\vec{r}_u \times \vec{r}_v|} \tag{15}$$

And $\vec{r}_u = \partial \vec{r}/\partial u$ and $\vec{r}_v = \partial \vec{r}/\partial v$ are used to simplify the notation. Equation (*14*) is equivalent to equation (7), precursor of equation (10) that describes the trajectory, but, in this case, to express the normal it is necessary to reconstruct the surface with the values of $\vec{r}$ near the considered ray. On the other hand, the direction of the normal has been chosen in the direction of advance of the front, that is, the same as $\nabla S$.

The combustion area, $A_b(t)$, at any given moment, is calculated by

$$A_b = \iint_{D(u,v)} |\vec{r}_u \times \vec{r}_v|\, du\, dv \tag{16}$$

traversing the set of parameters $D(u, v)$ that defines the combustion surface at each instant.

The temporal variation of the area is therefore

$$\frac{d}{dt}(A_b) = \iint_{D(u,v)} \frac{\partial |\vec{r}_u \times \vec{r}_v|}{\partial t} du\, dv \tag{17}$$

Differentiating the cross product yields

$$\frac{\partial}{\partial t}(\vec{r}_u \times \vec{r}_v) = \frac{\partial \vec{r}_u}{\partial t} \times \vec{r}_v + \vec{r}_u \times \frac{\partial \vec{r}_v}{\partial t} \tag{18}$$

The time derivatives of the position vector with respect to the parameters are obtained from equation (14):

$$\frac{\partial \vec{r}_u}{\partial t} = \frac{\partial \dot{r}_p}{\partial u}\vec{n} + \dot{r}_p \vec{n}_u \tag{19}$$

$$\frac{\partial \vec{r}_v}{\partial t} = \frac{\partial \dot{r}_p}{\partial v}\vec{n} + \dot{r}_p \vec{n}_v \tag{20}$$

Where the nomenclature is $\vec{n}_u = \partial \vec{n}/\partial u$ and $\vec{n}_v = \partial \vec{n}/\partial v$ for the derivatives of the normal vector. Substituting expressions (19) and (20) into (18),

$$\frac{\partial}{\partial t}(\vec{r}_u \times \vec{r}_v) = \dot{r}_p(\vec{n}_u \times \vec{r}_v - \vec{n}_v \times \vec{r}_u) - \left(\frac{\partial \dot{r}_p}{\partial u}\vec{r}_v - \frac{\partial \dot{r}_p}{\partial v}\vec{r}_u\right) \times \vec{n} \tag{21}$$

Note that the first term in the right-hand side of equation (*21*) is a vector perpendicular to the tangent plane (i.e. parallel to the normal direction) since both $\vec{n}_u$ and $\vec{n}_v$ are vectors contained in the tangent plane defined by $\vec{r}_u$ and $\vec{r}_v$. However, the second term is a vector perpendicular to the previous one, contained in the tangent plane.

Considering that $\vec{r}_u \times \vec{r}_v = |\vec{r}_u \times \vec{r}_v|\vec{n}$, it can also be written,

$$\frac{\partial}{\partial t}(\vec{r}_u \times \vec{r}_v) = \frac{\partial |\vec{r}_u \times \vec{r}_v|}{\partial t}\vec{n} + |\vec{r}_u \times \vec{r}_v|\frac{\partial \vec{n}}{\partial t} \tag{22}$$

and the comparison of equations (21) and (22) yields:

$$\frac{\partial |\vec{r}_u \times \vec{r}_v|}{\partial t} = \dot{r}_p(\vec{n}_u \times \vec{r}_v - \vec{n}_v \times \vec{r}_u) \cdot \vec{n} \tag{23}$$

$$\frac{\partial \vec{n}}{\partial t} = -\frac{1}{|\vec{r}_u \times \vec{r}_v|}\left(\frac{\partial \dot{r}_p}{\partial u}\vec{r}_v - \frac{\partial \dot{r}_p}{\partial v}\vec{r}_u\right) \times \vec{n} \tag{24}$$

Expression (23) evaluates the temporal evolution of the combustion area element, while expression (24) determines whether the direction of propagation changes or not, which is a result that had already been advanced, making use of the typical developments of geometric optics. These two expressions summarize the behavior of the combustion surface. If the recession rate is uniform, the direction of the normal at each point remains unchanged and the surface points move in a fixed direction. Conversely, if the recession rate changes from one point to another on the surface, the direction of the normal vector changes and the surface is distorted.

To further analyze expression (23), it is convenient to use some concepts of differential geometry of surfaces. The vectors $\vec{n}_u$ and $\vec{n}_v$ are contained in the tangent plane and can be expressed as a linear combination of the vectors $\vec{r}_u$ and $\vec{r}_v$, in the form

$$\vec{n}_u = a_{11}\vec{r}_u + a_{21}\vec{r}_v \tag{25}$$

$$\vec{n}_v = a_{12}\vec{r}_u + a_{22}\vec{r}_v \tag{26}$$

The matrix of coefficients is calculated by:

$$\begin{pmatrix} a_{11} & a_{21} \\ a_{12} & a_{22} \end{pmatrix} = -\begin{pmatrix} e & f \\ f & g \end{pmatrix}\begin{pmatrix} E & F \\ F & G \end{pmatrix}^{-1} \qquad (27)$$

where the coefficients of the First Fundamental Form (which corresponds to the inner product $d\vec{r} \cdot d\vec{r}$) are $E = \vec{r}_u \cdot \vec{r}_u$, $F = \vec{r}_u \cdot \vec{r}_v$, $G = \vec{r}_v \cdot \vec{r}_v$, and are related to the area element by $|\vec{r}_u \times \vec{r}_v| = \sqrt{EG - F^2}$. The coefficients of the Second Fundamental Form (which corresponds to the inner product $d\vec{r} \cdot d\vec{n}$) are $e = -\vec{n}_u \cdot \vec{r}_u = \vec{n} \cdot \vec{r}_{uu}$, $f = -\vec{n}_v \cdot \vec{r}_u = \vec{n} \cdot \vec{r}_{uv} = \vec{n} \cdot \vec{r}_{vu} = -\vec{n}_u \cdot \vec{r}_v$, $g = -\vec{n}_v \cdot \vec{r}_v = \vec{n} \cdot \vec{r}_{vv}$, and are related to the curvature of the surface.

The *normal curvature* of the surface is the ratio of both fundamental forms, $\kappa_n = (d\vec{r} \cdot d\vec{n})/(d\vec{r} \cdot d\vec{r})$, which is the component of the curvature vector $\vec{\kappa} = d\vec{t}/ds$ in the direction of the normal, where $\vec{t} = d\vec{r}/ds$ is the tangent vector (in this case the parameter $s$ describes any curve contained in $S$ that passes through the point in question). The normal curvature is independent of the curve on which it is defined and depends only on the orientation of the tangent vector. *Principal* curvatures are the maximum and minimum values of the normal curvatures of a given point. In particular, the main curvatures, $\kappa_1$ and $\kappa_2$, of the surface are the eigenvalues of the matrix $(a_{ij})$, the *average curvature*, $H = \frac{1}{2}(\kappa_1 + \kappa_2)$, is half of the trace of the matrix with the sign changed, $H = -\frac{1}{2}(a_{11} + a_{22})$, and *Gaussian curvature*, $K = \kappa_1 \kappa_2$, coincides with the determinant, $K = \det(a_{ij})$, which corresponds to the *intrinsic curvature* of the surface. Naturally, all these values do not depend on the chosen parameters.

Substituting expressions (25) and (26) into (23),

$$\frac{\partial |\vec{r}_u \times \vec{r}_v|}{\partial t} = \dot{r}_p(a_{11}\vec{r}_u \times \vec{r}_v - a_{22}\vec{r}_v \times \vec{r}_u) \cdot \vec{n} \qquad (28)$$

That is,

$$\frac{\partial |\vec{r}_u \times \vec{r}_v|}{\partial t} = -\dot{r}_p(\kappa_1 + \kappa_2)|\vec{r}_u \times \vec{r}_v| \qquad (29)$$

Expression (29), which can be rewritten as $d(\delta A)/dy = 2H(\delta A)$, is a classical result in differential geometry when one intends to obtain the variation of the area, $\delta A$, of a family of surfaces, and is directly related to very interesting topics, such as the plotting of surfaces of constant average curvature, or obtaining surfaces of minimum area. In the current context, it provides a direct geometric interpretation of how the combustion area evolves over time, due to the local value of the recession rate and as a function of surface curvatures. At a symmetrical saddle point, $\kappa_1 = -\kappa_2$, the net variation of the combustion area is zero, whereas, if the surface concavity prevails at the point, $\kappa_1 + \kappa_2 > 0$, the area decreases, but if the surface is globally convex, $\kappa_1 + \kappa_2 < 0$, the combustion area increases.

Similarly, equation (24) can be rewritten as

$$\frac{\partial \vec{n}}{\partial t} = -\frac{1}{|\vec{r}_u \times \vec{r}_v|^2}\left(\frac{\partial \dot{r}_p}{\partial u}\vec{r}_v \times (\vec{r}_u \times \vec{r}_v) - \frac{\partial \dot{r}_p}{\partial v}\vec{r}_u \times (\vec{r}_u \times \vec{r}_v)\right) \qquad (30)$$

The expression is apparently complicated, but if a new parameterization of the surface is used, being $(u', v')$ the arc lengths, it is then verified that $|\vec{r}_{u'}| = |\vec{r}_{v'}| = 1$ and if, in addition, orthogonality is required, i.e. $\vec{r}_{u'} \cdot \vec{r}_{v'} = 0$, this yields

$$\frac{\partial \vec{n}}{\partial t} = -\left(\frac{\partial \dot{r}_p}{\partial u'}\vec{r}_{u'} + \frac{\partial \dot{r}_p}{\partial v'}\vec{r}_{v'}\right) \equiv -\nabla_\parallel \dot{r}_p \qquad (31)$$

Where the vector identity $\vec{a} \times (\vec{b} \times \vec{c}) = \vec{b}(\vec{a} \cdot \vec{c}) - \vec{c}(\vec{a} \cdot \vec{b})$ has been used. To interpret the expression more easily one can use $dy = \dot{r}_p dt$ and write

$$\frac{\partial \vec{n}}{\partial y} = -\frac{\nabla_\parallel \dot{r}_p}{\dot{r}_p} \tag{32}$$

which is identical to (13). The normal vector to the surface changes its direction according to the direction marked by the gradient of the recession rate in the plane tangent to the surface and in the opposite direction. Equations (14) and (32) are a system equivalent to equation (10) that can be integrated over time, using the information provided by the function $\dot{r}_p(u,v,t)$, to obtain the evolution of the combustion surface,.

The normal vector is tangent to the trajectory followed by the point $P$, so its variation with the length traveled is the curvature, which will be proportional to the modulus of the gradient of the recession rate referred to itself, as written in equation (13). Admitting that this quantity is constant, for small values of the forward coordinate, the trajectory of this point describes an arc of radius $\dot{r}_p/|\nabla' \dot{r}_p|$.

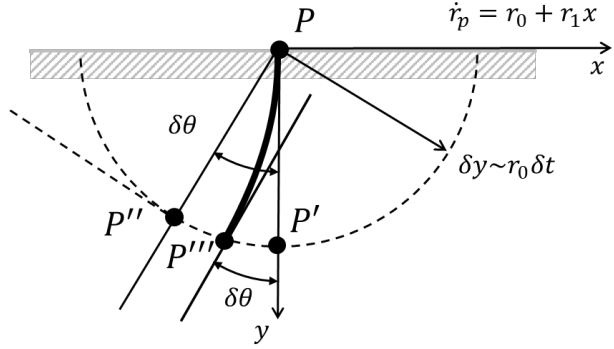

Figure 3: Schematic representation of the process that takes place when the recession rate varies linearly on the surface of the propellant. The trajectory of the ray is $PP'''$, although the point of tangency of the circle envelope is $P''$.

A schematic representation of the process that takes place with variable recession velocity has been made in Figure 3. The recession rate is considered to vary linearly on the surface of the propellant, $\dot{r}_p = r_0 + r_1 x$, and it is assumed that $r_1 \delta y \ll r_0$, so that the combustion front, for a time $\delta t$, moves from the point $P$ a distance $\delta y \sim r_0 \delta t$ that in the figure is used to draw a circle of center $P$ that locates the possible points that the combustion surface could reach. Applying Piobert's principle directly, equivalent to the exact result that the points on the surface move perpendicular to it, the ray would describe the trajectory $PP'$ and the new combustion surface would be built by joining the image points $P'$ of the entire surface. The above does not consider that applying Huygens' principle, each point $P$ of the surface is the center of a circle of distinct radii that grows at a rate $\delta\theta \sim r_1 \delta t$, being $P''$ the image points in a position other than $P'$. However, considering what was shown in previous lines, the trajectory of the point $P$ is an arc of radius $r_0/r_1$, and rotating an angle $\delta\theta \sim r_1 \delta t \sim r_1 \delta y / r_0$, to the point $P'''$. The points $P'$ and $P''$ are not correct and underestimate the position of the combustion surface, situation which is relieved because the distance $PP'''$ must be smaller.

### 3.1. Uniform recession rate

If the recession rate is uniform, then, $\partial \dot{r}_p / \partial u = \partial \dot{r}_p / \partial v = 0$ and from equation (30):

$$\frac{\partial \vec{n}}{\partial t} = 0 \tag{33}$$

Which indicates that the propagation directions remain unchanged, although the propagation velocity may be a function of time. These circumstances have the consequence of the propagation problem becoming decoupled from the temporal problem and being, therefore, purely geometric in nature. Consequently, the temporal evolution of the combustion area satisfies that the normal directions to the surface remain unchanged and the trajectories of the points on the surface are straight lines. The centers of curvature of the surface are located above the normal lines in fixed positions and the shape of the surface can be easily reconstructed. The surface retains its topology until the propellant consumption reaches some center of curvature. At that moment the analysis ceases to be valid and if the combustion front progresses there is an irreversible destruction of geometry.

Because $\dot{r}_p = dy/dt$, the forward coordinate may be used in equation (14), instead of time:

$$\frac{\partial \vec{r}}{\partial y} = \vec{n} \qquad (34)$$

Which is independent of the pace of recession and, therefore, the burnback problem is reduced to an exercise in geometry. Effectively, it can be integrated using the initial geometry from $y = 0$ (corresponding to the initial time, $t = 0$), obtaining a family of surfaces $\vec{r}_s(u, v, y)$. The recession rate can present any sort of time dependency because, from the known family $\vec{r}_s(u, v, y)$ and the expression $dy = \dot{r}_p(t)dt$, its evolution with time can be calculated.

Depending on the nature of the initial surface, different methods may be used to obtain its evolution. If the radii of curvature are defined at all points of interest, a possible procedure to obtain the surface $\vec{r}_s(u, v, y)$ is to evaluate the length of the radii of curvature, $\rho_{1,2} = 1/\kappa_{1,2}$, and describe how they change by means of equation (34). That is, solving $\partial \rho_{1,2}/\partial y = -1$, expression that supports the general solution

$$\rho_{1,2}(u, v, y) = \rho_{1,2}^o(u, v) - y \qquad (35)$$

Where the initial surface has the distribution $\rho_{1,2}^o(u, v)$ of radii of curvature. An immediate consequence is that, when a radius of curvature cancels out (note that, for this to be possible, it is necessary that the radius of curvature is strictly positive at $t = 0$, which corresponds to an initially convex geometry), there is an unavoidable discontinuity, since all the points of the combustion surface collide in the center of curvature, without the integration being able to continue. This event partially destroys geometry and requires a special analysis, since it is necessary to consider the evolution of a surface that contains non-regular points or regions.

Note that, contrary to the usual definition of the radius of curvature as the absolute value of the inverse of the curvature, here it has been given the sign of the curvature itself, to be able to generalize the relations. In this way, those radii that extend behind the space traveled by the normal are considered negative. When parameterizing the surface with the arc lengths, the absolute value of the radius of curvature will be taken, so that the angular sectors traveled will be positive.

Calling back to relationship (29):

$$\frac{\partial |\vec{r}_u \times \vec{r}_v|}{\partial y} = -\left(\frac{1}{\rho_1} + \frac{1}{\rho_2}\right) |\vec{r}_u \times \vec{r}_v| \qquad (36)$$

Where the time variable has been replaced by the normal coordinate. If the surface is parameterized by arc lengths following the main directions, $du' = |\rho_1|d\theta_1$ and $dv' = |\rho_2|d\theta_2$ (which are orthonormal, $|\vec{r}_{u'}| = |\vec{r}_{v'}| = 1$, when considering the main curvatures), the variation of the combustion area is:

$$\frac{dA_b}{dy} = -\iint_{D(\theta_{1,2})} \left(\frac{1}{\rho_1} + \frac{1}{\rho_2}\right) |\rho_1| |\rho_2| d\theta_1 d\theta_2 \qquad (37)$$

where $D(\theta_{1,2})$ expresses that integration variables now extend into a different domain than the parameterization used before. Note that the initial radii of curvature, $\rho_{1,2}^o = \rho_{1,2}^o(\theta_1, \theta_2)$, that we are going to use to calculate the area can be a function of the angles, $(\theta_1, \theta_2)$, so that there are no restrictions on the combustion surface, other than the mere existence of the radii of curvature. In the case of regions of null curvature, the original expression must be retrieved since the expression (37) has been invalidated by using the inverse of the curvatures in this analysis. Without going into major complications, what follows is useful to analyze the behavior of fixed angular sectors, since, for calculation purposes, the area can be decomposed into an arbitrary number of portions. With the intervention of (35) and some algebra, it can be obtained,

$$\frac{dA_b}{dy} = -\iint_{D(\theta_{1,2})} (\text{sgn}(\rho_1) |\rho_2^o - y| + \text{sgn}(\rho_2) |\rho_1^o - y|) \, d\theta_1 d\theta_2 \tag{38}$$

To analyze the expression, it is necessary to separate the different cases according to the sign of the curvatures, or the radii of curvature. If both are positive, the slope of the combustion area, depending on the forward coordinate, $dA_b/dy$, is monotonically decreasing and, as both radii decrease, the analysis is valid until the smaller one is canceled out. If both are negative, the slope is positive, with no limits other than those of the surface itself or those imposed by the combustion chamber casing. If the signs of the radii of curvature are different, it is necessary to elaborate the analysis with care. If we consider the case of the negative radius being less in absolute value than the positive one, the value of the initial slope is negative, and grows linearly with $y$ until it reaches the point where both radii equal in absolute value (this coincides with zero variation of the slope, which corresponds to a minimum of the local area enclosed in the angular sector considered). It then follows an upward slope behavior, until the initially positive radius is canceled out, stopping the linear analysis.

Any of the situations considered above leads to a linear variation of the slope and, therefore, to a quadratic variation of the combustion area with the forward coordinate of advance. Because of the above considerations, expression (38) can be reordered as follows:

$$\frac{dA_b}{dy} = \iint_{D(\theta_{1,2})} \text{sgn}(\rho_1 \rho_2) \left[2y - (\rho_1^o + \rho_2^o)\right] d\theta_1 d\theta_2 \tag{39}$$

The combustion area finally is

$$A_b = \iint_{D(\theta_{1,2})} \{\text{sgn}(\rho_1 \rho_2) [y^2 - (\rho_1^o + \rho_2^o)y] + \rho_1^o \rho_2^o\} d\theta_1 d\theta_2 \tag{40}$$

Which is canceled out when $y = \rho_{1,2}^o$ and, in addition, it is fulfilled $\rho_{1,2}^o > 0$, which corresponds to the situation of zero radius of curvature when the combustion front destroys a rounded cusp, already noted in the previous paragraph.

## 3.2. Cylindrical geometries

In line with the high slenderness of rocket-propelled aerospace vehicles, it is common to find combustion surfaces where the longitudinal dimension predominates. If the vehicle is very slender, and the thrust demand is high, the combustion surface must be greater than the cross-sectional area, and the only way to achieve this is by longitudinal drilling. In this case, the combustion surface is of cylindrical type, in which the characteristic dimension along the grain ($\sim L$) is large compared to the cross-sectional dimension ($\sim R$), that is, $L \gg R$. Local curvatures (in the longitudinal and transverse direction), necessarily, verify $\kappa_1 \sim 1/L$ and $\kappa_2 \sim 1/R$, which results in $\kappa_1 \ll \kappa_2$, so that equation (29) may be simplified by ignoring $\kappa_1$ as compared with $\kappa_2$:

$$\frac{\partial |\vec{r}_u \times \vec{r}_v|}{\partial t} \approx -\dot{r}_p \kappa_2 |\vec{r}_u \times \vec{r}_v| \tag{41}$$

The temporal variation of the combustion area is:

$$\frac{dA_b}{dt} = -\iint_D \dot{r}_p \kappa_2 dz ds \tag{42}$$

where the surface differential element, $|\vec{r}_u \times \vec{r}_v| du\, dv = dz ds$, is expressed by the coordinate along the cylinder $z$ and the arc length $s$. As the curvature of the cross section can be set as $k_2 = -d\varphi/ds$, being $\varphi$ the angle formed by the tangent to the curve, the previous expression becomes:

$$\frac{dA_b}{dt} = \iint_D \dot{r}_p dz d\varphi \tag{43}$$

For a differential element of area, it is verified:

$$\frac{d}{dy}(\delta A_b) = \delta z \delta \varphi \tag{44}$$

This is a very interesting expression. First, it shows again that if the recession rate is uniform then the problem is exclusively geometric. Moreover, if the length of the cylinder remains unchanged in the process then its influence is reduced to a constant factor. However, the most interesting property is that the variation of the combustion area *is independent of the shape of the cross section*. All reference to dimensions has disappeared from the expression. The variation of the combustion area is proportional to the value of the angular sector traveled by the tangent when running around the perimeter, and in the case of a straight cylinder of constant length ($L$) and uniform recession rate, it turns out to be:

$$\frac{dA_b}{dy} = 2\pi L \tag{45}$$

This value is independent of the shape of the section and corresponds to a progressive combustion process, identical to that which takes place for a cylinder of circular section. The combustion area is obtained immediately, $A_b = A_b^0 + 2\pi L y$, where $A_b^0$ is the value of the initial area for $y = 0$. Naturally, these results are subject to the cross-section being regular, in the sense that the curvature is defined at all points. Under these conditions, the variation of the area meets the following properties: *i*) it is independent of the shape of the perimeter; *ii*) it has a constant value equal to the angle rotated by the tangent to the curve; and *iii*) the sign (which marks the character of the combustion process) is the contrary to that of the curvature, when the normal to the curve points in the direction of propagation. Consequently, the expression of the perimeter is linear with the forward coordinate, and the process is reversible, in the sense that, if the direction of propagation is reversed, the initial geometry is reached uniquely. In the regressive regions of the perimeter, the propagation process decreases the radius of curvature and when the depth of advance reaches the center of curvature a discontinuity is generated, since a convex region disappears. At that point, the perimeter topology changes, and the surface analysis must be restarted, probably considering the evolution of a cusp, as will be discussed later.

The process of combustion of slender channels can be adequately described by one-dimensional models in which the geometry of the channel is determined by the distribution of port areas. Consider the perimeter of each section $P_b = \oint ds$ and the port area in each section $A_p = \oint r ds$. For calculation purposes, the combustion area, $A_b$, can be defined at any given section as the area of combustion exposed from $z = 0$ to the considered section $z$. That is,

$$A_b = \int_0^z P_b d\bar{z} \tag{46}$$

If the recession rate is uniform in the section, which is the most consistent simplification with the slender cylinder approximation, the variation with time of the port area (invoking again $\dot{r}_p = dy/dt = dr/dt$) is

$$\frac{dA_p}{dy} = P_b \qquad (47)$$

While the perimeter, in the assumption that it is a regular curve, complies with expression (45) and, therefore,

$$\frac{dP_b}{dy} = 2\pi \qquad (48)$$

The above expressions constitute a closed geometric system with which all geometric variables can be calculated using very simple integrals.

## 3.3. Non-regular geometries

The conclusions obtained in previous paragraphs can be generalized to contours in which the radius of curvature may present discontinuities, but for which the tangent to the perimeters must be a continuous function. In these circumstances, for each point of the combustion surface, an image point can be defined as the surface evolves. That is, a bijective relationship can be established between the points. This does not occur when: *i)* there are discontinuities in the tangent to the combustion surface, *ii)* two combustion surfaces meet each other, or *iii)* the combustion front reaches the motor case. In the first and second situations the trajectories of the surface points intersect. If the front reaches the motor case or any other inert element, the points on the surface also disappear irreversibly. All these situations are irreversible, in the sense that, if the sign of the recession rate is changed, the succession of combustion areas produced is not the same, just reversed in time, but very different, indeed. Next, a number of geometries that are commonly presented in solid propellant engines and that do not have a regular behavior are analyzed.

### 3.3.1 Corners and cusp

When the perimeter of the section presents a break, which represents a discontinuity in the slope, a non-regular situation is generated whose evolution is different depending on the direction of advance of the front. Figure 4 depicts two different situations in which the gaseous and solid domains are exchanged. In situation (a) the combustion process regularizes the geometry, the vertex of the corner becomes a source point, origin of a rarefaction, and as the geometry generated presents a smooth distribution of the angle (the tangent to the perimeter is continuous) the rate of increase of the perimeter, as already seen in equation (44), is

$$\left.\frac{dP_b}{dy}\right|_{Corner} = \Delta\varphi \qquad (49)$$

The increase (decreases are also possible) of the perimeter is proportional to the angle rotated by the tangent when following the curve. It is easy to imagine an algorithm that accumulates variations of the angle associated.

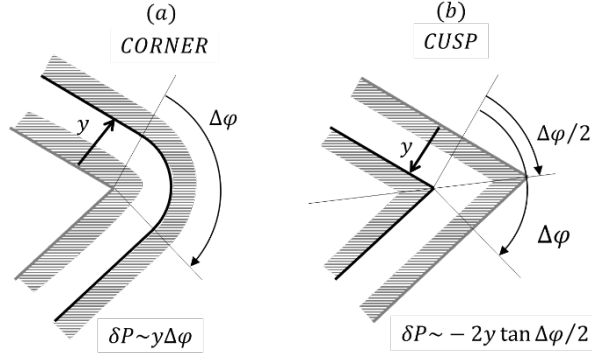

Figure 4: In configuration (a) the combustion front progresses from a corner creating a cylindrical surface (rarefaction). In configuration (b) the combustion front consumes a cusp destroying geometry and creating a discontinuity (caustic).

The situation (b) is the opposite to (a). The combustion front destroys part of the geometry as it advances. The collision of the two combustion fronts causes a discontinuity that is called a caustic. The destruction of geometry is irreversible. In Figure 4 (b), a simple geometric analysis leads to the relationship

$$\left.\frac{dP_b}{dy}\right|_{Cusp} = -2\,tan(\Delta\varphi/2) \qquad (50)$$

Which is similar to (49). Both relationships coincide if the angle rotated by the perimeter is very small ($\Delta\varphi \ll 1$) but, in general, equation (50) has a nonlinear dependence on the angle. Fortunately, both expressions have a linear dependence on the forward coordinate, and this allows combining different geometries so that any target value of $dP_b/dy$ can be set. Specifically, an adequate combination of valleys or corners (of a progressive nature, $dP_b/dy > 0$) and vertices or cusps (of a regressive nature, $dP_b/dy < 0$) can lead to a geometry in which the perimeter changes in a controlled way. For this case, the most common solution is a star-shaped geometry, in which the angle and number of cusps determines the progressive, regressive, or neutral character of the combustion.

### 3.3.2 Collisions

Figure 5 (a) shows a dendrite-like geometry in which, when the thickness is exhausted, the $2\omega$ arc of circle at the end of the protuberance disappears and the two combustion fronts collide simultaneously, producing an instantaneous drop (a discontinuity) of the combustion area of the form

$$\delta P = -\Delta P_D \mathcal{H}(y - \omega) \qquad (51)$$

where, $\delta P$ is the discrete variation of the perimeter, $\Delta P_D$ is the length of the dendrite, $\mathcal{H}(\,)$ is the Heaviside function and $\omega$ is the semi-thickness of the dendrite (remember that $\mathcal{H}(x) = 0, x < 0$; and $\mathcal{H}(x) = 1, x \geq 0$).

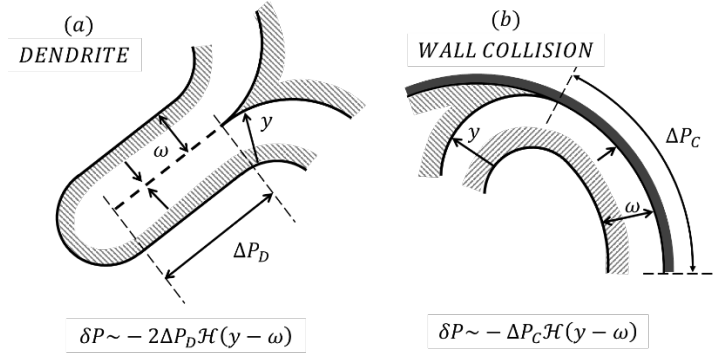

Figure 5: Sometimes there is the collision of two combustion front (a) or with the engine casing (b), which causes a sudden destruction of the combustion surface and a discontinuity in the evolution of the perimeter.

A similar situation occurs when cylindrical combustion surfaces collide with the motor case in the final phase of propellant combustion. If the collision takes place sharing the center of curvature the decrease in area will be sudden,

$$\delta P = -\Delta P_c \mathcal{H}(y - \omega) \tag{52}$$

Being here $\Delta P_c$ the arc length of the collision front. If the collision is not completely frontal, a very rapid process of combustion area destruction occurs, which must be analyzed in each case.

In both these situations, the most relevant characteristic is that the processes are not linear. In fact, the burning area presents discontinuity that originate an unsteady response of the chamber pressure. Furthermore, the evolution is not reversible.

# 4. Burnback analysis methods

Current methods can be classified into Analytical or Numerical. The analytical methods, essentially, consist in using Piobert's aphorism and displacing the combustion surface, formed by simple geometric figures, perpendicular to itself, incorporating the particular phenomenology imposed by cusps and corners. This activity can be carried out for a simple geometry obtaining closed relationships, or by automating operations through some algorithm, such as the SPP© program or other CAD-type graphic programs. In contrast, numerical methods start from a discrete description of the combustion surface, which allows them to be more flexible and general. Once the discretized surface is available, it can act as in analytical methods using some specific property of the solution or address the propagation problem by integrating differential relationships.

| Method | | References |
|---|---|---|
| 1. Analytical methods | | |
|   1.1. Simple/unique geometry | | [2]–[7],[8], [9] |
|   1.2. Combination of simple geometries | | [1], [10]–[12], [13] |
|   1.3. CAD based methods | | |
|     1.3.1. Parametrized geometry | | [14]–[16] |
|     1.3.2. Based in CAD in-house tools | | [17]–[19] |
| 2. Numerical methods | | |
|   2.1. Direct surface tracking | | [20]–[23] |
|   2.2. Minimum distance function (MDF) | | [24]–[26], [27]–[30] |
|   2.3. Theory of curve and surface evolution (PDE's based) | | |
|     2.3.1. Set Level Methods (Hamilton-Jacobi equation) | | |
|       2.3.1.1. Standard (signed function evolution) | | [34]–[41],[42]–[47] |
|       2.3.1.2. Narrow band | | |
|     2.3.2. Steady perspective (Eikonal equation) | | |
|       2.3.2.1. Direct time marching | | [48]–[53] |
|       2.3.2.2. Fast marching methods (FMM) | | [54] |

Table 1: Classification of the different methods of burnback analysis.

Table 1 lists all the categories of methods considered in this paper and indicates the most relevant bibliographic sources. Applied to geometries accessible to the method, all those listed in the table solve the problem satisfactorily, from the point of view of thrust curve calculations. Numerical methods are usually able to deal with more general and complex problems than analytical methods, though. Different arguments have been raised in the literature to evaluate the suitability of each method. As the most versatile and powerful methods are numerical methods, the central argument is usually efficiency, measured in terms of computational time requirements. However, the high power achieved by computers today weakens the importance of this argument, because the computational effort in the field of burnback analysis is moderate compared with that required for the study of, for example, the rocket internal aerodynamics or the structural calculation of the propellant. Numerical burnback analyses only need to obtain a single spatial function that determines the combustion surface as the propellant is consumed. In addition, it is not necessary to use adapted meshes, but with significantly uniform meshes that reasonably describe the geometry is sufficient to obtain satisfactory results. From this perspective, other considerations, such as the flexibility in terms of the possibility of carrying out complex three-dimensional geometric analyses, the possibility of analyzing cases with variable recession velocity, and the economy of implementation, all make the methods based on the Eikonal equation (2.3.2 in Table 1) the most attractive. This result contrasts with the very high diffusion that LSM have reached in the analysis of the burnback problem in the last twenty years, motivated by the evident generality of the method. However, the burnback analysis problem does not need so much generality, and the LSM is oversized in this case. The integration of the Eikonal equation is enough to obtain a completely satisfactory solution of the problem (that is, the calculation of the thrust curve) and, eventually, allow the design of the initial combustion geometry. In the next section, the results obtained with method 2.3.2.1 of Table 1, which meets the above requirements, are presented for a variety of grain geometries.

## 4.1. Analytical methods

Analytical methods make use of different properties of the solution that are incorporated into the analytical calculation of the position of the combustion surface. These methods are fast, simple and accurate. But they cannot address problems of arbitrary geometry, they have to solve complex geometric situations with specially adapted procedures, and they cannot, in general, solve problems of variable recession velocity.

### 4.1.1 Simple/unique geometry

It is the simplest approach and consists in the algebraic analysis of the evolution of a surface that moves perpendicular to itself. During the second third of the twentieth century, in the early days of the development of solid propellant engines, it was the only possible method. The work of Billheimer and Wagner [2] contains an extensive bibliographical review of this period and the different procedures with which the simple geometric calculation was enriched to achieve the determination of the thrust curve of solid propellant engines with grain geometries that presented some complexity. For example, the work of Thibodaux et al. [4] can be reviewed to verify the level of specialization achieved in the analysis of, in this case, three-dimensional geometries in spherical chambers. Or the arduous work of analyzing the interaction between the combustion front of a slotted-tube grain with the casing of the engine, described in [5]. This type of method is still widely used, and the number of recent citations, referring to burnback analysis with purely analytical methods, is very high (not all collected in this review), because the immediacy of the method lends itself to its easy integration into internal aerodynamics analysis systems [5], or its integration into all kinds of engine design optimization algorithms [6][7].

As already mentioned, the most interesting advantages of the method are its speed, simplicity, and precision. Naturally, it is not possible to analyze arbitrary geometries and it is difficult to incorporate realistic situations such as a non-constant recession velocity. In addition, the analyses must incorporate a specific treatment of non-continuous geometries (such as cusps and corners) which, for example, in three dimensions can significantly complicate the problem. However, it is possible to address situations of industrial interest and others that initially would seem complex, such as the analysis of two propellants burning simultaneously. In this sense, through analytical methods, it is possible to address the problem of two propellants with two different recession rates, as for example, to analyze the combustion of a bipropellant star geometry that does not present sliver mass fraction[3][8] and that Krishnan and Bose [9] study with a high level of detail for various configurations.

### 4.1.2 Combination of simple geometries

The simplicity of use of analytical methods facilitates a different strategy, combining elements of simple geometry and automating the analysis of the evolution of the combustion surface. The best known and most successful example is the burnback module of the SPP© software package, initially presented by Coats et al. [1] in 1987, and continuously updated and improved since then (see [11]–[13]).

The SPP© program has been a standard reference software in the United States for predicting the performance of solid-propellant rocket engines. The methodology for evaluating the thrust coefficient, starting from the chemical equilibrium value, which is corrected with individual efficiencies due to different effects, is an industry standard. The Grain Design and Ballistics module allows the design of the initial combustion surface and calculates the thrust curve using a burnback analysis package, an internal aerodynamics module, and calculations with finite chemical kinetics in a two-dimensional nozzle flow. The SPP© program has been used in the past, and is still being used today, by major agencies, institutions, and manufacturers of solid-propellant rocket engines in the United States and other countries [13]. The grain design and analysis module construct the surface by extracting simple geometric figures from an initial volume (the interior of the motor case). It is a Boolean operation that can be repeated with the basic figures resized. In the calculation of the evolution of the combustion surface, the dimensions of geometric figures are increased, emulating the advance of the front. Operationally, the program is fed with symbolic commands, which are executed sequentially. It is a flexible, versatile, and efficient tool, capable of modeling all the geometries that are usually presented in solid propellant rocket engines, as long as they can be decomposed into simple volumes. Naturally, it is an analytical methodology that retains the disadvantages already mentioned, but the product has been adapted and consolidated to mitigate these disadvantages as much as possible.

**4.1.3 CAD based methods.**

The increase in accessibility and power of computer-aided design (CAD) has meant that these specialized programs have been used to conduct burnback analysis of realistic and overly complex geometries. This is the main quality of the method, the ability to evaluate surfaces of complicated shapes. Two strategies may be adopted for the calculation of the area evolution.

On the one hand, when modeling the initial combustion surface, the model can be parameterized so that the recession process is considered, using the parameters that define the model itself (1.3.1. in Table 1, see references [15]–[17]). For example, if a cylinder is parameterized by its radius, by varying the radius a preset quantity, the process of recession is simulated. The next operation is to vary the parameterized values and allow the graphic system to reconstruct the new combustion surface, executing the corresponding symbolic operations.

The other possibility is to use CAD-specific capabilities that move the model surface with controlled laws (1.3.2. in Table 1, see references [17]–[19]). That is, specific tools for translation, growth, or projection of surfaces that the software makes available to the user. These procedures are quick and versatile, can tackle complex geometries, and provide a fast and adequate response. However, the information obtained must be extracted from within the CAD system. Furthermore, there is no certainty of these geometric operations being able to capture the real problem physics, since many of these operations are hidden from the user. Naturally, the user is forced to examine these operations and, eventually, correct situations in which the graphical system fails because it is unable to automatically generate rarefactions or caustics.

## 4.2. Numerical methods

Numerical methods approach the problem from a discrete description of a combustion surface. Depending on the method, the initial combustion surface can be an external surface of a volumetric mesh that represents the whole propellant, where other surfaces of interest can be easily identified as well, such as the motor case or, for example, symmetries of the model. Alternatively, the combustion surface is discretized as an isolated surface, whose movement is the objective of the calculation and which, in one way or another, must incorporate an analysis of the interaction with other surfaces such as the engine casing. The advantage of numerical methods is that they allow the description of the evolution of complex combustion surfaces, and, with some exceptions, they allow variable recession rate to be incorporated into the calculation.

**4.2.1 Direct surface tracking**

This category includes methods that carry out local surface monitoring, combined with a position identification that allows interaction with inert areas or with the engine casing. In principle, this type of methods start from a discretization of the surface and obtain its evolution using displacement algorithms that somehow consider properties exhibited by the propagation process. The most commonly used of these properties is Piobert's postulate that the surface moves perpendicular to itself. Typical methods of calculating free surfaces (Volume of fluid, VOF, method) are also used, identifying the convection rate with the recession rate of the front.

Among these methods, one can mention the SLIC (Simple Line Interface Calculation) method devised by Noh and Woodward [20]. The authors conceived it for use in one, two or three spatial dimensions. The domain is discretized into enclosures, and fluid interfaces are represented locally for each enclosure by lines, either perpendicular or parallel to the coordinate directions. Decision-making logic is used in the propagation, depending on the arrangement of fluid regions. Due to the completely one-dimensional nature of the interface description in SLIC, it is relatively easy to get correct results with time. Another very similar method is FLAIR [55], which tries to increase the accuracy by complicating a little the geometric description of the front within each control zone, and is used by Mashayek et al. [21] for the analysis of two-dimensional combustion geometries.

Belonging to the surface tracking methods that use phenomenological algorithms, which basically project the surface perpendicular to itself, the work of Hejl and Heister [22] carries out direct surface tracking and incorporates locally the peculiarities that are presented in the form of rarefactions and caustics. Also, in reference [56],[56][57],[57]Another work in this category is carried out by Ki et al. [23] that present the PIT method (Partial Interface Tracking) in the analysis of combustion surfaces of three-dimensional geometrics of type finocyl and conocyl. This method applies a Lagrangian approach to the axisymmetric area of the transverse plane and the two-dimensional area of the longitudinal plane separately, because the Lagrangian approach is an effective way to simulate two-dimensional evolutions. In this way, a three-dimensional problem is solved with the computational effort of two two-dimensional problems. The limitation is that geometries have to exhibit some symmetry, which is usually common in solid propellant engines, such as finocyl and conocyl types. However, it does not bring anything new in the spectrum of front-tracking methods, but it merely solves with success three-dimensional problems approximately.

### 4.2.2 Minimum distance function (MDF)

The method of calculating the minimum distance to the initial combustion surface, proposed by Wilcox et al. [24], has been very fruitful in solving the burnback analysis problem and, in this case, is used to allow internal ballistic calculation [25]. It is a very intuitive method, easy to implement, and does not exhibit limitations in terms of the complexity of the geometry. Once the domain occupied by the propellant has been discretized, the method consists in calculating the smallest distance from any interior point to the initial surface. This calculation involves a search for the point of the initial surface closest to the inner point, which is onerous from the computational standpoint. Usually, methods of reducing this computational time are required, optimizing search algorithms using standard techniques, such as Ren et al. [26] using a divide-and-conquer algorithm. Like other methods already discussed, MDF employs a property of the solution, in this case, Fermat's Principle, and when the propagation velocity is uniform, the minimum time condition is equivalent to the minimum distance condition. Precisely, this is the disadvantage of the method, which cannot incorporate variable recession rate without overcomplicating the algorithm. The reason the generalization of the MDF method is not possible is that a global property is used, which leaves out of the calculation what is the path followed by each ray. However, the conceptual simplicity and the possibility of applying it in realistic three-dimensional geometries, makes it a widely used method [28]–[30], see, for example, how in [31] is concluded that it is superior to other surface monitoring methods.

### 4.2.3 Theory of curve and surface evolution (PDE's based)

To describe the propagation of the combustion surface in solid propellant rockets, Saintout et al. [48] implement an algorithm that incorporates all the characteristics that allow it to describe the physics of the process properly, and identify the equation that they integrate numerically as of the Hamilton-Jacobi type. This situation is reached from preliminary studies of the same research group on surface tracking methods [50] and [49]. These works are part of the activity made by SNPE (Société Nationale des Poudres et Explosifs, currently a subsidiary of Nexter) for the analysis and design of solid propellant rocket engines in Europe that, in the case of burnback analysis, culminate with the work of Dauch and Ribereau [51]. In this work, the general purpose tool called PIBAL© is presented, which integrates an evolution of the IVOLINA© program (previously developed in the references [52] and [53]) that addresses the integration of the Eikonal equation by a time marching method (2.3.2.2 in Table 1).

However, in parallel to the developments described in the previous paragraph, for the treatment of this type of problems (and other, more complex), the work of Osher and Sethian [31] initiates a lineage of methods, based on the procedure called Level Set method (LSM). These methods have been very fruitful and have been developed and employed on numerous occasions (see [32] for an overview). In what follows, it is described how the problem has been solved by two different paths, the first addresses the resolution of an equation of type Hamilton-Jacobi by means of the LSM that is capable of solving problems of propagation of very general fronts, much more complex than the problem of burnback. The second perspective addresses the steady problem that is circumscribed to the solution of an equation of type Eikonal that is strictly the problem to be solved in the burnback analysis and, in this sense, the modeling and computational effort made is more proportionate. A basic and complete description of both approaches can be obtained in Sethian's text [33], which clearly identifies and discusses both methods.

Consider the situation in Figure 6 in which a curve or surface, defined for example by the function $\phi = 0$, spreads with velocity $\dot{r}_p$ in the direction perpendicular to the surface itself. The problem is to determine the evolution of the surface. In the most general situation, the propagation rate may depend on local properties of the surface point, such as the direction of the normal, or curvature, or on general properties of the curve, such as integral relations of all kinds, and, also, on properties external to the problem itself, as would be the case of advection, due to a velocity field.

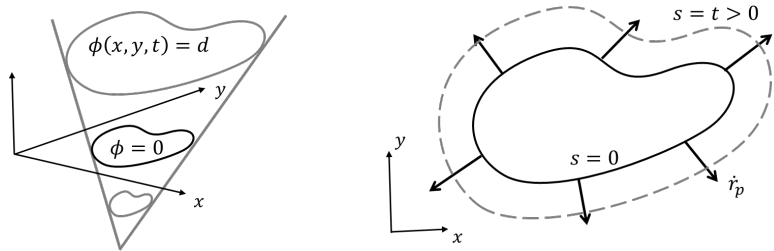

Figure 6: Outline of the two approaches followed in the numerical methods of solving the burnback problem. On the left, the LSM in which the front is represented by the null value of a distance function, $\phi = 0$. On the right, the position of the front is represented by the values taken by the solution of the Eikonal equation that corresponds to the travel time.

The problem can be approached from two points of view. The boundary value formulation calculates the time $\tau = s(x)$ it takes for the front to reach each point in the domain, and it is evident that the definition of the velocity of the front leads to $\dot{r}_p = \delta x/\delta \tau$ and, therefore, in several dimensions it is fulfilled that,

$$\dot{r}_p |\nabla s| = 1 \tag{53}$$

Already written before, with the condition $s = 0$ on the initial combustion surface. This is the Eikonal equation, which is a traditional problem in many physical systems. In this problem, it has been implicitly assumed that the function $\tau$ is a single-value function, for which the propagation rate must have a constant sign, either outwards from the domain, or inwards. This restriction, which for some situations is very important, in the case of burnback analysis is fulfilled naturally and the unknown of the Eikonal equation is strictly the function to be obtained to solve the evolution of the combustion surface in a solid propellant rocket.

When propagation can take place in two directions, on both sides of the front, it is mandatory to describe the movement of the front by a function $\phi$ with more spatial dimensions. To obtain an equation of this evolution, consider the path $\vec{x}(t)$ that follows a particle of the front and how, without loss of generality, one can assume the front defined by $\phi(\vec{x}(t), t) = 0$. Differentiating the function yields,

$$\phi_t + \nabla \phi(\vec{x}(t), t) \cdot \vec{x}'(t) = 0 \qquad (54)$$

Which is the equation that allows us to obtain the function $\phi$. As the velocity of the front is $\dot{r}_p = \vec{n} \cdot \vec{x}'(t)$ and the direction normal to the surface is $\vec{n} = \nabla \phi / |\nabla \phi|$, finally, the equation for $\phi$ is:

$$\phi_t + \dot{r}_p |\nabla \phi| = 0 \qquad (55)$$

For which an initial value of the function must be supplied. This equation is of the Hamilton–Jacobi type, for a wide spectrum of forms of $\dot{r}_p$. The problem of front propagation occurs in a wide variety of configurations: from ocean waves, combustion fronts or interfaces in the movement of heterogeneous substances; of course, in problems of light propagation or seismic wave propagation; but also, in problems of character identification or image processing.

Equations (53) and (55) represent the two different approaches, and both of them provide fully satisfactory results. The only difference is that solving the Eikonal equation involves an effort adjusted to the problem. The method based on the Hamilton-Jacobi equation is designed for more complex problems and needs further elaboration in the calculation, uses more memory and has to solve numerical problems (such as the reinitialization of the distance function) typical of a more complex method, but which are totally unnecessary in the burnback analysis problem.

4.2.3.1. Level Set Methods (Hamilton-Jacobi equation)

The evident generality of LSM has led the methodology to be used in the analysis of the burnback problem on numerous occasion [34-47]. Usually, the initial function $\phi(\vec{x}, t = 0)$ is fixed as a signed distance function (SDF) containing the value of the minimum distance to the front from the initial surface and which is calculated with some algorithm (2.3.1.1 in Table 1). The method is not exempt from some problems, since the SDF can take poorly conditioned values as the integration progresses, and it becomes necessary to reinitialize it periodically. [58]

It is evident that the method employs an implicit function defined throughout the propellant domain of which the only useful information is the front defined by the null value of the function. For this reason, some authors have used a strategy of limiting the calculated value of the SDF to the vicinity of the front (2.3.1.2. in Table 1). However, it is necessary to incorporate a search and location algorithm of the front to determine the narrow band.

Notable is the contribution of Chiapolino et al. [47] that addresses the solution of a Hamilton-Jacobi equation using a standard LSM but with a step function for the level function emulating the front tracking methods commonly used in heterogeneous fluid problems. An instructive article describes a numerical method on an unstructured mesh, in which it uses upwind techniques with limiters, for the method stability, which have been developed in previous works. The examples that are included, addressing three-dimensional burnback analysis, are very illustrative, and correspond to modern and realistic grain geometries.

4.2.3.2. Steady perspective (Eikonal equation)

4.2.3.2.1. Direct time marching

The solution of equation (53) (also of the equation (6) using a Time Marching procedure),

$$s_t + H(\nabla s) = 0 \tag{56}$$

where the Hamiltonian is

$$H(\nabla s) = 1 - \dot{r}_p |\nabla s| \tag{57}$$

As already mentioned, this type of equation belongs to the so-called *Hamilton-Jacobi equations*, which arises as a problem of initial values with boundary conditions according to the situation to be simulated. In the problem in hand, $s = 0$ on the initial combustion surface. Usually, the rest of the boundary conditions consist of boundaries at which the front extinguishes (for example, the engine casing) in which, usually, no condition is necessary to be imposed due to the hyperbolic nature of the equation; and contours of symmetry or periodicity, in which the implementation of the condition is relatively simple.

References [48]–[53] pioneer the use of the Eikonal equation for the solution of the burnback problem. These works constitute a frame of reference for the correct and adjusted solution. Since then, however, the propagation problem has been addressed from different perspectives and for different problems, though not necessarily in solid-propellant engine technology. The direct solution of the Eikonal has been addressed on numerous occasions for the appropriate monitoring of surfaces, as in [59]. Singular is the contribution of Gueyffier et al. [60], which addresses the solution of the Eikonal equation using a spectral method for the description of the combustion surface with a philosophy similar to that employed by surface tracking methods.

4.2.3.2.2. Fast marching methods (FMM)

The Eikonal equation in the form (53) can be solved by calling a method based on the traditional alternate direction methods but using the propagation direction of the front to update the variables and in this way obtain an additional advantage. These procedures are called Fast Marching methods (FMM). It is possible to consult the book by S. Sethian [54] to have an overview, where an interesting critical comparison between FMM and LSM is also established. The burnback problem has been addressed by this method in unstructured mesh, for example, in [61] in a complete paper but there are not many other contributions to the burnback problem using this procedure.

# 5. Burnback analytical solutions

Constant combustion surface area is the most common design condition for a solid propellant rocket engine. This situation is generically optimal because it implies that the chamber structural design is adjusted to the entire engine operation range. Otherwise, the thickness of the engine casing must be sized for the most unfavorable load case, which corresponds to the maximum pressure reached and, therefore, the combustion chamber is heavier than that of the engine that would provide the same total impulse with constant chamber pressure. To achieve constant pressure profiles with large combustion areas, comparable to those of the chamber itself, it is necessary to resort to geometries with a certain degree of complexity. The important variables are the web fraction, the volume fraction, and the sliver fraction but, also, the Klemmung and $J$ (combustion to port area ratio). These last two parameter are of interest because control the occurrence of the erosive combustion phenomenon.

## 5.1. Classic star

Figure 7 shows the geometric description of the cross-section of a star-shaped propellant with $n$ tips, as discussed in [62], [63]. For simplicity, only half of an angular sector, $\pi/n$, is represented, taking advantage of the symmetry properties of the section. The tip has an angle $\theta$ (the figure shows the semi-angle $\theta/2$) while occupying a fraction $\varepsilon$ of the entire angular sector, $\varepsilon(\pi/n)$. The depth of the valley area has length $d$ from the center of the chamber and it is considered that the thickness of the propellant is the necessary to finish the first phase of combustion at the moment in which the front arrives for the first time to the engine casing. If the propellant web thickness is larger, a progressive phase of linear perimeter growth begins at this time. This second phase of combustion would be progressive, and in the design of the engine it will not be allowed to extend too much, as it raises the chamber pressure. However, it can increase the web fraction or the combustion time, and it may be necessary to satisfy design requirements. Nevertheless, the possibility of compensating this effect by designing the star with a slightly regressive profile should be analyzed.

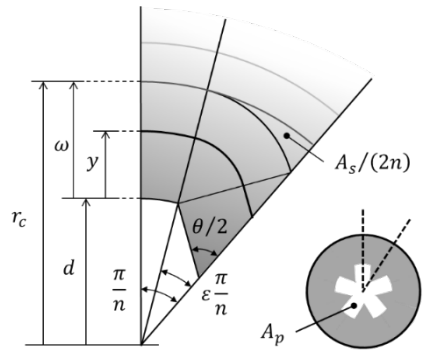

Figure 7: Schematic of star geometry and definition of geometric parameters.

A heuristic procedure to determine the variation of the perimeter of the section considered is to go through the contour, measuring the rotation suffered by the normal to the surface and calculating, in each case, the increase in perimeter that occurs. Performing this operation for the geometry in Figure 7, the expression obtained is

$$\Delta P = \underbrace{2n}_{(1)} y \left\{ \underbrace{\frac{\pi}{n}}_{(2)} + \underbrace{\left(\frac{\pi}{2} - \frac{\theta}{2}\right)}_{(3)} - \underbrace{\frac{1}{tan[\theta/2]}}_{(4)} \right\} \qquad (58)$$

The term (1) corresponds to the $2n$ half sectors, the term (2) is due to the turn suffered by the normal in the half sector (if it were a cylinder, these first two factors would give rise to the simple result already commented $\Delta P = 2\pi y$). The term (3) is the one corresponding to orienting the normal from the radial position, after the rotation $\pi/n$, to the surface of the cusp, which assumes a rotation equal to the complementary angle of $\theta/2$. Finally, the term (4) is the one corresponding to the destruction of part of the cusp. Indeed, the first thing to note is that $\theta/2$ is the complementary angle of the angle $\Delta\varphi/2$ in the Figure 4, and the tangent function of the complementary angle is the inverse of the tangent of the angle and, in addition, in the generic expression (equation (50)) two slopes of the cusp are taken into account, while in the sector in Figure 7 only one of them has to be accounted for.

This rapid assessment is delicate and subject to probable misinterpretation of the criteria under which it is applied. However, it is a very interesting method to be used in combination with more elaborate geometric evaluations. Because it provides a quick verification. In addition, it allows us to carry out analyses that lead to the elaboration of optimal strategies for the design, merely using analytical arguments.

To obtain a geometry in which the combustion area does not change (i.e. neutral combustion), it is necessary that $\Delta P = 0$ in equation (58) which is a simple nonlinear equation for $\theta$ as a function of $n$. Table 2 shows the semi-angle of the tip, the web fraction, and the volumetric fraction, for different values of the number of tips. Note that between 5 and 6 tips, the value of $\theta$ goes from being $\theta/2 < \pi/n$ to be $\theta/2 > \pi/n$, showing that for less than 5 tips impossible geometries can arise in which the tips collide with each other. It is especially interesting that if $\theta/2 = \pi/n$ the channel is straight. This is the condition for analyzing axial slots. The table indicates as well that for $n \geq 6$ (because $\theta/2 > \pi/n$) the combustion process should be regressive, which is very useful information when combining combustion geometries.

| $n$ | 4 | 5 | 6 | 7 | 8 |
|---|---|---|---|---|---|
| $\theta/2$ | 28.21 | 31.12 | 33.53 | 35.55 | 37.30 |
| $\pi/n$ | 45.00 | 36.00 | 30.00 | 25.70 | 22.50 |
| $\omega/D_c$ | 0.200 | 0.180 | 0.164 | 0.151 | 0.140 |
| $\chi$ | - | 0.893 | 0.804 | 0.733 | 0.674 |

Table 2: Solution of equation (58) for neutral combustion $\Delta P = 0$, and the corresponding values of the web fraction and the volumetric fraction.

Combined propellant geometries are presented on many solid rockets. A common configuration is to use simple cylindrical combustion and a slotted segment. The cylindrical section has a combustion area that grows over time and the slotted segment can be configured so that the combustion area decreases at the desired rate. The combination of both geometries can result in a thrust curve with a specific profile.

### 5.1.5 Bipropellant star

The star configuration provides a constant combustion area curve for moderate values of the web fraction. However, the mass of residual propellant after the neutral phase (*sliver* fraction) can be very large, with a negative impact on the effective volumetric fraction. It is possible to design a sliverless geometry using two propellants with different recession velocities. The idea is to fill the region of the cusp with a high-speed recession propellant so that it reaches the engine casing at the same time as the propellant, with a lower recession rate, that fills the web thickness.

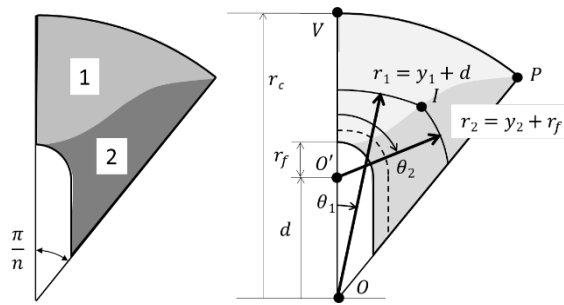

Figure 8: Straight star loaded with two propellants of different recession rates. On the left is a general scheme and on the right the notation used in the analysis to determine the adequate interface.

Figure 8 shows a possible simple configuration for a straight star (similar to a slotted geometry) in which two propellants of different combustion rate are used. Each propellant advances a different amount at the same time due to the different rate of combustion. The propellant 1 advances $y_1$ and the propellant 2 advances $y_2$. For the combustion front to reach the casing simultaneously at all points, the combustion front in the propellant 1 has to be cylindrical with radius $r_1$ from point $O$. While, by construction, the combustion front in the propellant 2 is composed of a line and a circle arc of radius $r_2$, centered on point $O'$. The recession velocity in the propellant 2 must be such that it reaches the point $P$ at the same time as the propellant 1. This imposes a geometric exception, depending on which combustion front reaches the point $P$ in the propellant 2, whether it is the straight front or the circular front. In what follows it is assumed that it is the circular combustion front that reaches the point $P$. The rest of the parameters to be used are shown in Figure 8, in which $d$ is the depth of the slot, $r_f$ is the fillet radius in the slot, $r_c$ is the chamber radius and $r_{1,2}$ and $\theta_{1,2}$ are the polar coordinates of the points on the two combustion fronts, respectively. The condition for the fronts to progress simultaneously over the interface is expressed by

$$r_1 \cos\theta_1 = r_2 \cos\theta_2 + d \tag{59}$$

$$r_1 \sin\theta_1 = r_2 \sin\theta_2 \tag{60}$$

Where $r_1 = r_f + d + y_1$ and $r_2 = r_f + y_2$. Without loss of generality, it can be put $y_1 = y$ and $y_2 = fy$ with $f > 1$. The regression rate of the propellant 2 is suitable so that, on the symmetry line, the front reaches the housing at the point $V$ at the same time as in the propellant 1 reaches point $P$. The propellant 2 induces a cylindrical combustion front on the propellant 1 with radius $r_1$, while the front in the propellant 2 is also cylindrical with radius $r_2$ but with center at $O'$. The condition of reaching the casing simultaneously at the point $P$ is expressed by removing $\theta_2$ from expressions (59) and (60), thus getting

$$(r_1 \cos\theta_1 - d)^2 + (r_1 \sin\theta_1)^2 = r_2^2 \tag{61}$$

And substituting $r_1 = r_c$, $\theta_1 = \pi/n$ and $y = \omega$, which is the value of the web thickness, result in:

$$r_c^2 - 2r_c d \cos(\pi/n) + d^2 = (r_f + f\omega)^2 \tag{62}$$

Along with

$$r_c = r_f + d + \omega \tag{63}$$

Once the geometry of the star is established ($n$, $r_c$, $r_f$ and $d$ are known), equation (63) allows the calculation of the web and equation (62) provides the needed value of $f$, that is, the ratio of recession velocities of both propellants, $f = r_{p2}/r_{p1}$.

The geometry of the interface can be obtained by taking as a parameter the depth of the forward coordinate of propellant 1 ($y; 0 \leq y \leq \omega$) and explicitly resolving with

$$r_1 = r_f + d + y \tag{64}$$

$$r_2 = r_f + y_2 \tag{65}$$

And using equation (61) to obtain $\theta_1$ and equation (60) to obtain $\theta_2$. Once the interface line has been drawn, it is possible to calculate the burn perimeter on each propellant and, considering the different recession velocities, calculate the mass released by each propellant. The length of each perimeter in each propellant is no longer an intuitive measure of the mass burned by the entire surface or of the chamber pressure reached at each moment. For this reason, in what follows, the geometric concept of forward coordinate is momentarily abandoned in favor of a pseudotime, as an independent variable.

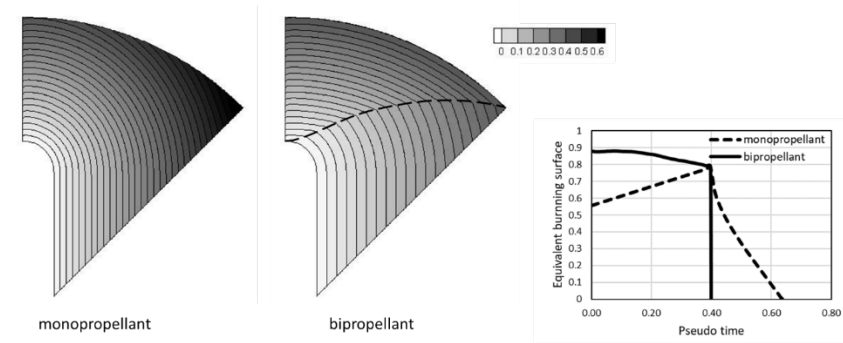

Figure 9: Comparison of the equivalent burning surfaces of a monopropellant and a bipropellant star-shaped geometry with 4 cusps, with $\omega = 0.4$, $r_f = 0.1$ and $d = 0.5$; so that the ratio of recession velocities takes the value $f = 1.592$.

Figure 9 shows the simulation performed with a geometry of four slots. The results obtained in the case of operating with a single propellant and in the case of operating with two sliverless propellants are presented. To compare both situations, it is useful to represent the pseudotime lines that correspond to $\tau = y_{1,2}/r_{p1,2}$. Taking $r_{p1} = 1$ in the case of a single propellant, the pseudotime is equivalent to the forward coordinate. In the bipropellant case, an equivalent combustion area must be defined in the form $A_{b,eq} = A_{b,1} + fA_{b,2}$. The equivalent combustion area allows the calculation of the mass released and the chamber pressure and thrust, using the combustion data of the propellant 1. In this way, we can establish a reliable comparison with the operation of a single propellant. In the monopropellant case, this geometry, with few slots, gives rise to an increasing combustion area profile, until the combustion front reaches the engine casing for the first time. From that moment, the combustion area decreases over time, giving rise to a long tail thrust phase as shown in the figure. In the bipropellant case, however, the combustion process of the fast propellant generates at the beginning more mass flow, compensating the initial deficit presented by the monopropellant. In this way, as clearly shown in Figure 9, a near-neutral combustion area curve is provided. This remarkable feature can be anticipated by designing the geometry so that the equivalent combustion areas are similar at the initial and final times. As the combustion fronts reach the casing simultaneously, no sliver fraction is produced and the combustion area curve drops sharply at that moment, forming an optimal silverless geometry.

Figure 10 shows the comparison of operation with one and two propellants of an elliptical hole geometry that initially presents a high volumetric filling. As a result, the ratio of recession rates is also high, which translates, again, into a significant variation in the equivalent area of combustion. In this case, the pseudotimes at the end of the combustion of both configurations are equal, highlighting the significant variation (up to 33%) in the equivalent combustion area. For the calculation of the interface, the approximation of the combustion front in propellant 1 remaining elliptical has been made. The numerical simulation shows how little importance this gross hypothesis has on the overall result.

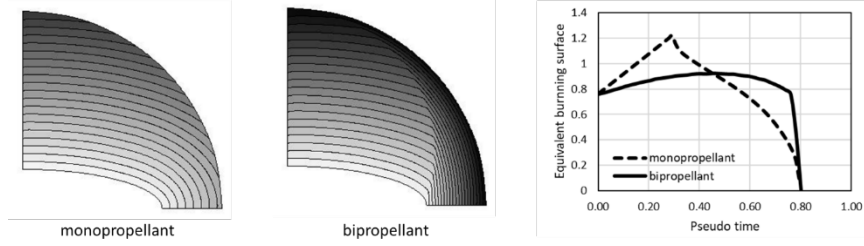

Figure 10: Comparison of the equivalent burning surfaces of a monopropellant and a bipropellant elliptical-hole geometry, with high volumetric fraction and high ratio of recession velocities $f$.

The solution of these bipropellant cases has been approached without establishing any consideration about how the combustion fronts interact with each other. As will be seen below, the interaction can be complex and create rarefaction and caustic waves that significantly modify the combustion front near the interface. This can lead to variations of some importance in the evaluation of the equivalent combustion area and, therefore, in the prediction of the actions of the system. As will also be shown later, the numerical analysis scheme proposed reliably captures these anomalies. For the examples presented above, this anomaly does not occur, since it is a corner-type combustion situation in which the design system guarantees that the interface is above the equilibrium point $E$ (the scheme $i$)) in Figure 14, so that the combustion fronts do not present rarefactions or caustics of any kind.

## 5.2. Bipropellant burnback analysis

The combustion front in a bipropellant grain is determined by the difference between the recession rates of each propellant and by the geometry of the front and of the interface. To approach a general analysis with confidence, it is advisable to start with a simple situation, in which the combustion front at the point of contact of both propellants is flat, as represented in Figure 11. The point $S$ separates both propellants at the combustion surface, and the interface between them is straight and perpendicular to said combustion surface. The recession rate of the propellant 1 (on the left in the figure) is $r_1$ and the combustion front of this propellant moves to the parallel line $b_1$ a distance $y_1 = r_1 \delta t$ after a time $\delta t$, at points that are far enough from the point S. At the same time, the combustion front for the propellant 2 moves $y_2 = r_2 \delta t$ reaching the line $b_2$. For the analysis it will be assumed that $r_1 < r_2$ and, therefore, $y_1 < y_2$. To build the solution it is convenient to consider the point $S$ to be the source of the propagation process in both propellants, then the combustion front will extend into the propellant 1, at least, up to the cylinder $c_1$; and, in the propellant 2, up to the cylinder $c_2$. As in time $\delta t$ propellant 2 reaches the point $F_2$, while propellant 1 only would reach point $F_1$, faster propellant acts as a source of ignition. Each of the points on the side of the propellant 2 over the line $\overline{SF_2}$ will be the center of a family of circles that consumes the propellant 1, and whose radius is proportional to the distance remaining to travel to $F_2$. Consequently, the combustion surface in the propellant 1 will be the envelope of this family of cylinders, which is easily built by tracing the tangent to the circle $c_1$, from $F_2$ to the point of tangency $T$. The segment $\overline{F_2T}$ intersect with line $b_1$ at the point $C$, separating the combustion surfaces obtained from the original surface ($b_1$), and that obtained because of the phenomenon already described in the interface ($\overline{F_2C}$). On point $C$ two different combustion fronts converge, whose collision forms the caustic $c$, which is a straight line starting from the point $S$ in the line $\overline{SC}$.

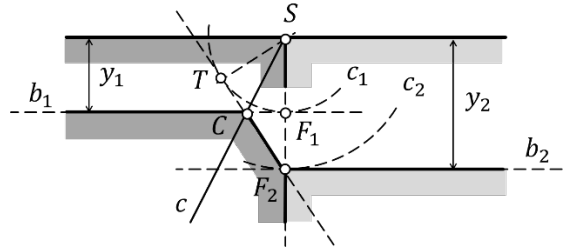

Figure 11: Diagram of the evolution of the combustion surface of a bipropellant with flat front and interface perpendicular to the front.

The propagation of a combustion front, initially flat, along the interface between two propellants is a relatively common situation that, for example, corresponds to that which occurs in the case of thermally conductive wires, embedded in the propellant to increase the combustion area. In this case the cable acts as an ignition source with a higher velocity than the propellant regression rate and the combustion geometry obtained is conical with the axis on the cable, analogous to the construction $\widehat{b_1CF_2}$ in Figure 11. However, although the situation is simple, it allows the introduction of the basic analysis mechanisms to be used in more complex situations. Thinking that the point S It is the origin of the combustion front of each propellant, building the cylinders of influence $c_1$ y $c_2$, calculating the intersection with the lines that establish the position of the fronts $b_1$ and $b_2$ far from $S$, and determining the envelope of certain families of cylinders, leads to the construction of the combustion surface at each time.

For the analysis of more complex situations, where the combustion front is not initially flat, it is convenient to generalize the notation, as shown in Figure 12. Uppercase letters are used to name points of interest and lowercase letters are used to name lines and circular arcs. The figure represents the two possible situations for a non-flat combustion front, when $S$ is the vertex of a cusp, and when it is the vertex of a corner. The bisector angle $\beta$ is used to represent the initial position of the fronts and identify the angle $\delta$ (that lies between the lines $e$ and $f_2$) as a measure of the difference in burning rates, because if $\delta = \beta$ the burning rates are equal.

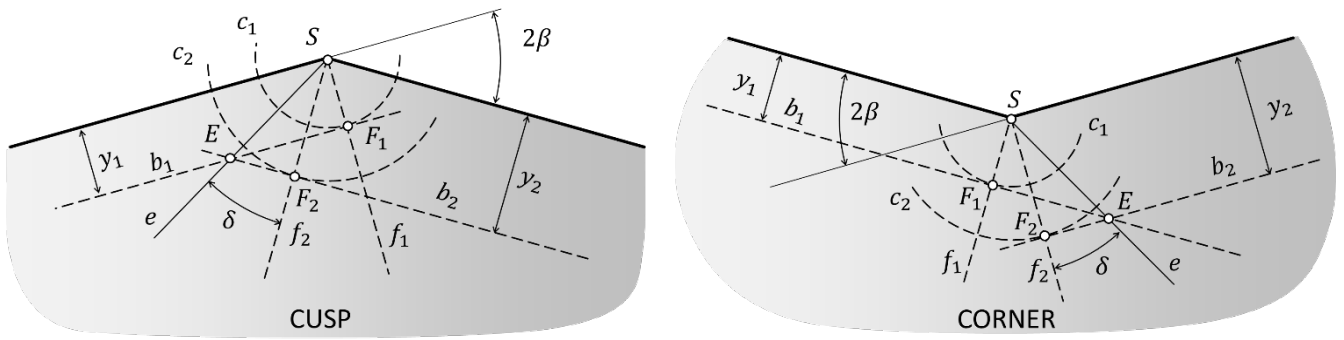

Figure 12: Meaning of the different symbols used in the description of the propagation process of a bipropellant for two initial configurations of the initial combustion surface.

The lines $b_1$ and $b_2$ are parallel to the original surfaces and represent the position of each combustion front if they were isolated (in the figure, $y_1 < y_2$). Unlike the simple flat-front case, lines $b_1$ and $b_2$ are not parallel, but rather converge at the point $E$, that allows you to draw the equilibrium line $e$ from $S$, towards $\overline{SE}$. Once the position of the surfaces is known, the circles of influence, $c_1$ and $c_2$, can be traced, tangents to the aforementioned lines at points $F_1$ and $F_2$, respectively. The perpendicular to the initial surfaces, $f_1$ and $f_2$, are drawn from $S$ following the directions $\overline{SF_1}$ y $\overline{SF_2}$. The conical zone between $f_1$ and $f_2$ defines an interference region. The equilibrium line $e$ is a reference for the position of the interface between the propellants, which will evolve differently if it is inside or outside the interference region. Finally, if the angle formed by the initial surfaces of both propellants is $2\beta$, then the angle of the equilibrium line $e$ can be calculated through the relationship:

$$\frac{y_1}{y_2} = cos(2\beta) - sin(2\beta)\, tan(\delta) \tag{66}$$

which shows that the structure of the study region depends on $\beta$ and the recession rate ratio, that is, $y_1/y_2$.

Figure 13 shows the different results of the combustion surface if $S$ is the vertex of a cusp. Each scheme corresponds to distinct positions of the interface, identified by the line $i$, determined by angle $\delta_i$ that the interface forms with the line $f_1$ (Figure 12). The series starts with a sufficiently large value of the angle $\delta_i$ (greater than $\delta$) and situations are analyzed for decreasing values of $\delta_i$. If $\delta_i > \delta$, see diagram $a)$, the interface intercepts the line $b_1$ at the point $I$ and the propagation process in the propellant 1 produces the premature ignition of propellant 2 along the interface between them. The combustion surface produced (segment $\overline{IC}$) is generated by obtaining the line that starts from $I$ and is tangent to the cylinder $c_2$, which corresponds to the envelope of the family of circles generated by the ignition points. The intersection of this line with the line $b_2$ determines the position of the point $C$ which is the vertex of caustic $c$ generated in this process which goes from $S$ towards $\overline{SC}$. As $\delta_i$ decreases, the point $C$ approaches point $I$, coinciding both when $\delta_i = \delta$, and the caustic disappears, as illustrated in the diagram $b)$ of Figure 13. In this situation, the propellants are consumed at their own rate without generating any additional structure, forming the fronts only by the lines $b_1$ and $b_2$. When $\delta_i < \delta$, but before you get to $\delta_B$, the propellant 1 reaches point $I$ before propellant 2, but this time the family of cylinders ends in the circle that passes through $I$ and is tangent to $b_2$ and, as represented in the scheme $c)$, a partial rarefaction is created between lines $f$ and $i$. When $\delta_i$ reaches the value of $\delta_B$ (scheme $d)$) the expansion is complete between the line $f_2$ and the line $i$. Here, the point B, that defines $\delta_B$, is obtained as the intersection of the circle $c_2$ and the line $b_1$. Until now, the high inclination of the interface line causes the process to be dominated by the propellant 1 but when $\delta_B > \delta_i$ the propellant 2 reaches the point $I$ earlier, producing the premature ignition of the propellant 1. In the scheme $e)$ this situation is shown, in which the new surface $\overline{IC}$ is obtained by drawing the line that starts from $I$ and is tangent to $c_1$, obtaining the position of $C$ as an intersection of this line and $b_1$. Under this, if $\delta_i > 0$ remains the rarefaction between $i$ and $f_2$ until $\delta_i$ is canceled (scheme $f)$) and rarefaction disappears. Finally, as represented in the scheme $g)$ for $\delta_i < 0$ the structure of the front is maintained.

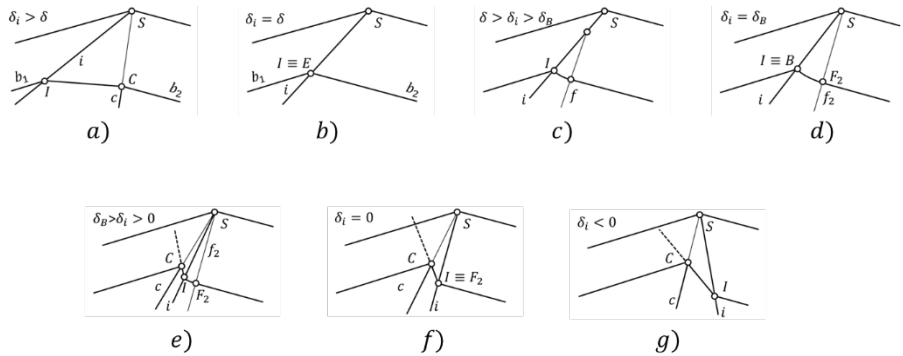

Figure 13: Sequence of the different schemes for different positions of the interface, in the case where the initial combustion surface has a cusp at the vertex between the two propellants.

For the case where in cusp configuration the point $E$ is within the interference region (delimited by $f_1$ and $f_2$) the different modes of propagation of the Figure 13 are simplified and only schemas $a)$, $b)$ and $g)$ appear.

Figure 14 shows the different morphologies of the combustion surface when $S$ is the vertex of a corner separating both propellants. Each scheme corresponds to distinct positions of the interface, identified by the line $i$, determined by angle $\delta_i$ that the interface forms with the line $f_2$ (Figure 12). When $\delta_i$ is large enough, as depicted in the scheme $h)$, the propellant 2 runs through the entire interface to the point $I$ faster than propellant 2. Therefore, the combustion surface $IT$ is plotted by calculating the envelope of the propagation cylinders in the propellant 1, that is, it is obtained from the line that starts from $I$ and is tangent to $c_1$ at the point $T$. As depicted in the scheme, between the lines $f_1$ and $f$, a rarefaction is formed that is reduced as $\delta_i$ decreases. The rarefaction disappears when $\delta_i = \delta$ (scheme $i)$) which, as in the cusp, gives rise to a scenario without mutual interactions, each propellant was consumed independently of the other. While $0 < \delta_i < \delta$ the process is, as presented in the scheme $j)$, similar to the scheme $h)$, but in this case the point of tangency $T$ goes over the line $f_1$ and the envelope generates the caustic $c$ along the segment $\overline{SC}$.

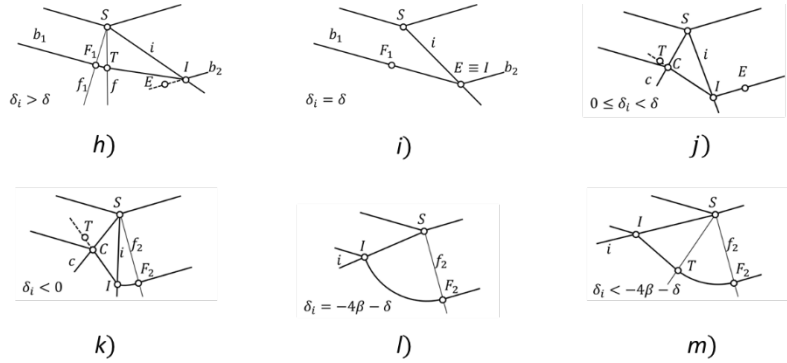

Figure 14: Sequence of the different schemes for distinct positions of the interface, in the case where the initial combustion surface has a corner at the vertex between the two propellants.

The situation when $\delta_i < 0$ is represented in scheme $k)$ of Figure 14. A rarefaction is generated in the propellant 2 between the lines $i$ and $f_2$. The surface of the front coincides with $b_2$ up to $F_2$, and between $F_2$ and $I$ it coincides with $c_2$. As the propellant 2 continues to dominate the process, the combustion surface $\overline{IC}$ is obtained as before, tracing the tangent to the circle $c_1$ that goes through $I$. For negative angles ($\delta_i < 0$), but greater in absolute value, the circular arc $F_2I$ grows until it reaches the line $b_1$ in which, as represented in scheme $l)$, caustic $c$ disappears, when reaching the interface itself. If the interface tilts even more (scheme $m)$) is now the propellant 1 the one that dominates the propagation process, causing the ignition of the propellant 2. The envelope $IT$ is created from the cylinder $c_2$.

All the above situations correspond to the scheme of Figure 12 at which the equilibrium point $E$ is outside the interference region bounded by $f_1$ and $f_2$. When point $E$ is situated within the interference region, schemes $h)$, $l)$ and $m)$ are reproduced and a new configuration, not represented in the figures, appears with two rarefactions, one in each propellant next to the lines $f_1$ and $f_2$.

Figure 15 shows the result of numerical analysis, with a code based on obtaining the solution of the Eikonal equation, by simple time marching, which is described later in this work. As can be seen in the figure, the structures of the schemes $g)$, $j)$ and $k)$ are reproduced faithfully. The algorithm efficiently captures rarefaction and caustic structures produced near the interface between the two propellants.

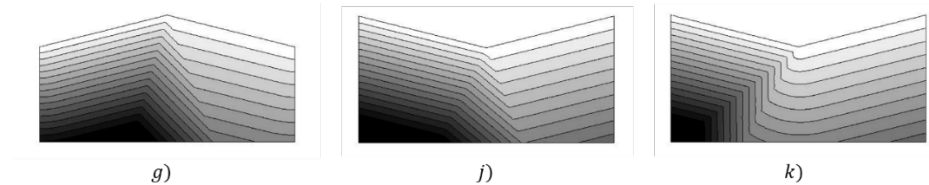

Figure 15: Result of the numerical simulation of three examples of bipropellant interface corresponding to the same reference letters in Figure 13 and Figure 14.

On the one hand, this shows that the previous analytical reasoning is correct, in general terms. On the other, it shows the power and versatility of the numerical method proposed. As has been seen in the review of the literature and the analysis of the different methods, the numerical integration of the Eikonal equation is the best procedure to establish the combustion surfaces, even when the recession rate is variable. Using a discrete representation and computation system, the kinematics of geometrically complex burning fronts propagating with prescribed variable burning rates can be efficiently described.

# 6. Burnback numerical solution

As shown in the previous sections, the solution of the burnback problem by numerical methods that offers the best results passes, in the general case, through the solution of a Hamilton-Jacobi equation, although, naturally, the direct solution of the Eikonal equation can be selected. Two lines of work can be distinguished in the solution of this type of equations, one that addresses the mathematical problem in a generic way (e.g. [64]–[66]) and another driven by the solution of front propagation problems using LSM (e.g. [67] or, very recently, [47]).

The study of numerical approximations to the viscous solution was also initiated by Crandall and Lions [65]. They introduced an important class of monotonic schemes for a simplified form of equations and showed that these schemes converge to the viscous solution (for an in-depth review of this matter from a general point of view, see [64]). However, it is known that monotonous schemes can be at most first-order, so they are too dissipative for most practical applications, although they are used to build high-order algorithms. In reference [67], Osher and Sethian built a class of high-order upwind-type schemes to, imitating ENO algorithm of high order developed by Harten et al. [68] and Shu and Osher [69], approximate conservation laws. Its construction was based on the observation that the Hamilton-Jacobi equations are closely related to conservation laws. In this sense, a wide variety of algorithms have been proposed, such as those described in e.g. [69]–[71].

In particular, in the problem of burnback analysis, this type of algorithms has been used on numerous occasions, but the applications that are most interesting are those developed for unstructured meshes. The nature of the initial combustion surfaces and the need to use complicated geometries that meet the design requirements of solid-propellant rocket engines leads inexorably to the use of unstructured meshes. In addition, this type of meshing allows noticeably short generation times, which has a significant impact on the overall efficiency of the process. The solution of the unstructured Hamilton–Jacobi equation composed of triangles was first proposed by Abgrall [72] by the approximate solution of a classical Riemman problem, based on the work of Bardi and Evans [73]. These works have been followed by others [74]–[77] in which the approximation order was increased or different schemes of the same type were tested. Special mention should be made, in this category, of the schemes that obtain the solution of the Eikonal equation by means of fast marching algorithms in unstructured meshes, such as in [78] or [79].

## 6.1. Time marching method

In the present work, the solution of the Eikonal equation is obtained by means of the simple time marching procedure in an unstructured two-dimensional mesh composed of triangular elements. The integration domain is the complete volume of propellant, delimited by the initial combustion surface and the surfaces that remain inert (surfaces inhibited for combustion and the surfaces in contact with insulating material or in contact with the case). The value of the unknown function $s(\vec{r}, t)$, which represents the time of arrival of the front, is stored at the vertices of the mesh and, as already indicated above, the problem to be solved is

$$s_t + H(\nabla s) = 0 \tag{67}$$

In which the Hamiltonian is

$$H(\nabla s) = 1 - \dot{r}_p |\nabla s| \tag{68}$$

With the initial condition $s(\vec{r}, 0) = 0$, which is also imposed as a boundary condition on the initial propellant surface throughout the integration. The method used does not need to impose spatial boundary conditions on inert surfaces, through which the combustion front passes without disturbance. However, it is customary to select portions of the propellant volume delimited by surfaces with symmetry conditions, which is easily implemented in the algorithm.

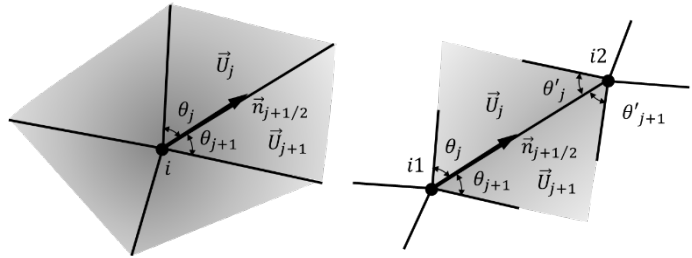

Figure 16: In the diagram on the left, the main geometric elements used in the basic discretization around the node $i$ are represented; and on the right, the notation used in the edge-based algorithm to construct the discrete solution is shown.

The solution of the equation can be obtained numerically efficiently, by means of a discretization based on the work of Abgrall [72]. This requires a domain triangularization, using the variable values $s_i$ $(i = 1..nn)$ at the vertices, to estimate the value of the gradients of the function at each triangle, $\vec{U}_j = [\nabla s]_j$ $(j = 1..nt)$. In Figure 16, the geometric configuration used is represented, in which the angles around an edge connected to the node i are $\theta_j$ y $\theta_{j+1}$ and the unit vector in the direction of the edge is $\vec{n}_{j+1/2}$. The value of the function over time $t = (n+1)\Delta t$ is obtained from:

$$s_i^{n+1} = s_i^n + \Delta t \mathcal{H}(\vec{U}_j^n) \tag{69}$$

Where

$$\mathcal{H}(\vec{U}_j^n) = H\left(\frac{1}{2\pi}\sum_j \theta_j \vec{U}_j^n\right) - \varepsilon_i \sum_j \beta_{j+1/2} \frac{\vec{U}_j^n + \vec{U}_{j+1}^n}{2} \cdot \vec{n}_{j+1/2} \tag{70}$$

And

$$\beta_{j+1/2} = tan\frac{\theta_j}{2} + tan\frac{\theta_{j+1}}{2} \qquad (71)$$

Integration must be carried out under the stability condition $\Delta t \leq h/L$ where $h$ is the minimum height of the adjacent triangles and the diffusion factor is calculated by $\varepsilon_i = L/\pi$, being $L = \max_j \|\nabla u\|$.

The algorithm is constructed by traversing the edges of the mesh and updating the value of the mean gradient on each node (see the right diagram of Figure 16). The procedure is executed by using the following relationships:

$$\vec{U}_{i1} \leftarrow \vec{U}_{i1} + \frac{1}{2}(\theta_j \vec{U}_j + \theta_{j+1} \vec{U}_{j+1}) \qquad (72)$$

$$\vec{U}_{i2} \leftarrow \vec{U}_{i2} + \frac{1}{2}(\theta'_{j+1} \vec{U}_{j+1} + \theta'_j \vec{U}_j) \qquad (73)$$

And calculating

$$\beta_{j+1/2} = tan\frac{\theta_j}{2} + tan\frac{\theta_{j+1}}{2} \qquad (74)$$

$$\beta'_{j+1/2} = tan\frac{\theta'_{j+1}}{2} + tan\frac{\theta'_j}{2} \qquad (75)$$

The diffusion terms of the equation are calculated by

$$D_{i1} \leftarrow D_{i1} - \varepsilon_{i1}\beta_{j+1/2}\frac{1}{2}(\vec{U}_j + \vec{U}_{j+1}) \cdot \vec{n}_{j+1/2} \qquad (76)$$

$$D_{i2} \leftarrow D_{i2} - \varepsilon_{i2}\beta'_{j+1/2}\frac{1}{2}(\vec{U}_{j+1} + \vec{U}_j) \cdot (-\vec{n}_{j+1/2}) \qquad (77)$$

Boundary conditions are applied for the nodes of each contour by modifying the values of $\vec{U}_i$ and of $D_i$ calculated on all nodes as follows:

a) Free contour

$$H_i \leftarrow H_i \qquad (78)$$
$$D_i \leftarrow 2D_i \qquad (79)$$

b) Symmetry contour

$$\vec{U}_i \leftarrow \frac{1}{2}(\vec{U}_i + \vec{U}_i|^{sim}) \qquad (80)$$
$$D_i \leftarrow 2D_i \qquad (81)$$

Where $\vec{U}_i|^{sim}$ is the symmetric vector to $\vec{U}_i$. The integration is advanced until reaching a steady state, which is ensured by checking that the gradients of the variable within each triangle do not change above a predetermined value.

## 6.2. Results and discussion

Previously, throughout this document, results of numerical simulations that employ the algorithm described above have been presented in Figure 9, Figure 10 and Figure 15. These results clearly show the method's ability to deal with situations in which the velocity of front propagation is not constant. The cases of bipropellant, in star configuration and ellipse of high fill coefficient, are handled efficiently. The presence of the interface that separates both propellants is undertaken with the single implementation of assigning different values to the recession rate to each node within the domain. The same technique is used in the three simulations presented in Figure 15, configuring the calculation domain and the inclination of the interface properly.

Figure 17 shows the results of three representative cases. The results have been calculated with unit recession rate in the system of units in which the geometry is represented, that is, advance coordinate and pseudotime coincide. The three cases have been calculated with a modest number of elements not exceeding $10^4$ nodes. Even so, the results show reasonable precision in the absence of a more rigorous error analysis that is carried out in the following section. The examples show the ability of the method to describe all relevant phenomena in the analysis of these configurations. In the so-called anchor geometry, the combustion front collides with the engine casing generating an abrupt change in the combustion area, while, inside, a caustic is formed when the combustion fronts collide, coming from the central slot and the circumferential groove. The second case corresponds to a star geometry optimized to produce neutral combustion. Finally, an unoptimized case of dogbone geometry is included in which it is observed that the condition of free contour in inert boundaries is treated without visible reflections and disturbances.

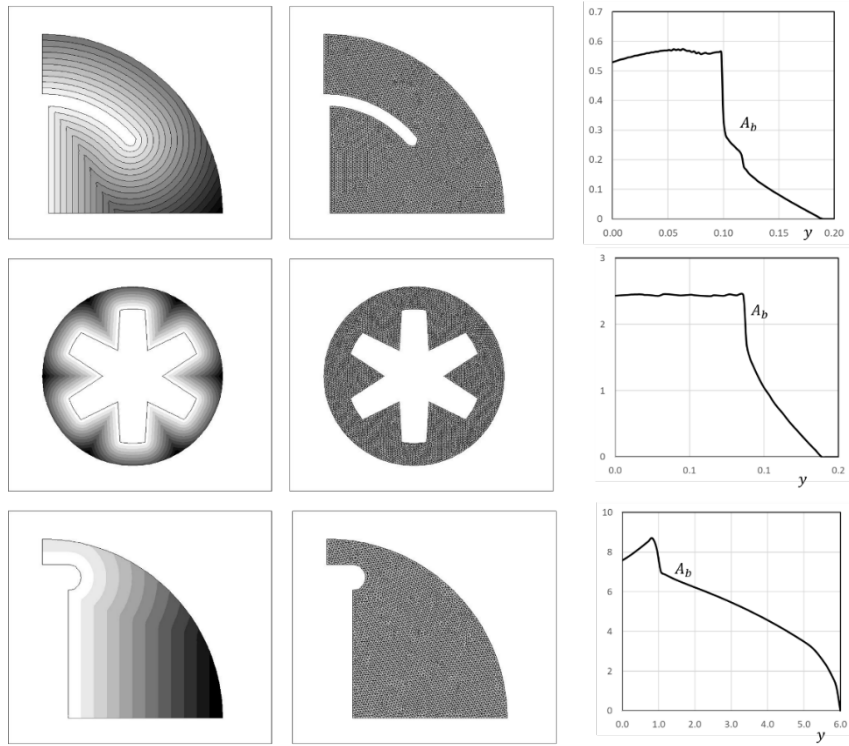

Figure 17: From left to right: constant pseudotime lines, mesh utilized, and curves of combustion surface area for three representative cases (top to bottom: anchor geometry, optimized neutral-burn star, and unoptimized dogbone).

Figure 18 shows the results obtained with a partially optimized axil-geometry. The constant pseudotime line shows the full variety of situations; and the combustion area curve only needs a few adjustments to present a properly flat profile.

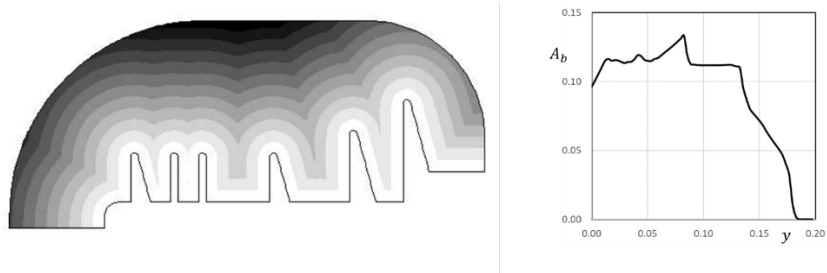

Figure 18: Constant pseudotime lines and combustion area curve for a geometry corresponding to a low-slenderness engine with an axil-type grain in the process of manual optimization.

**6.2.7 Error analysis**

A simple slotted geometry is chosen to perform error analysis. This geometry brings together two aspects of interest: the expansion of a combustion front in which the combustion perimeter increases and the collision of two combustion fronts with the consequent generation of a caustic. This is a simple situation, and the error can be calculated by comparing it with the analytical solution of the problem.

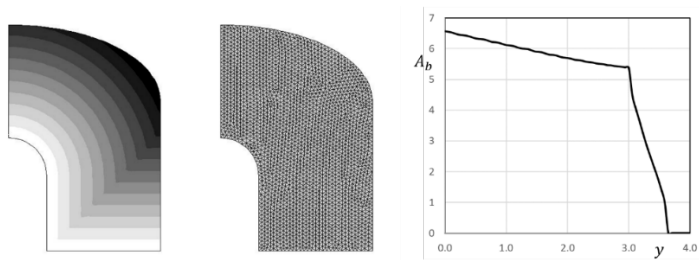

Figure 19 Level contours of pseudotime (left), 2500-node mesh (center), and combustion area (right) on the problem used to evaluate the discretization error.

The problem consists in the advance of a combustion front from a radial slot (only the right half of the domain is considered using vertical symmetry) composed of a straight section ending in a semicircle. As shown in Figure 19, the motor case would be located at the upper border and at the right border where the combustion front leaves the domain. The combustion area that develops this geometry is traced in the graph of the figure, and consists of a first section of neutral combustion, due to the increase in perimeter caused by the circular expansion, combined with the destruction of geometry caused by the caustic, followed by a process of a strong fall of the area, while the combustion front leaves the asymmetrical upper part.

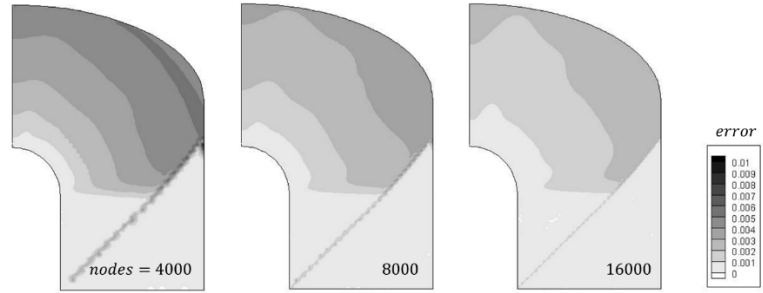

Figure 20: Constant normalized error contours in the single slot problem for different number of nodes in the mesh.

Figure 20 displays the normalized error obtained in simulations with different number of nodes in the mesh. The error has been calculated as the difference between the calculated value and the exact analytically calculated one. The error is normalized with the maximum value of the penetration level reached by the front, so the contours of the figure are representative of the relative error. This procedure has been chosen because it is not possible to calculate the relative error in the initial contours in which the value of the forward coordinate is very small. In all cases, it is observed that the error incurred is extremely small in the advance of the straight fronts. However, on the discontinuity the error is noticeable and in absolute value increases throughout the expansion range. In the cases analyzed, the maximum error, corresponding to the coarsest mesh, is less than 1% and is located where the discontinuity crosses the contour. By increasing the number of nodes of the mesh, the error decreases significantly and as already mentioned, even with meshes of modest size, the results obtained are very valuable. In the figure, the denser mesh provides a solution in which the error in the front position is less than 0.3%.

# 7. Conclusions

Burnback analysis is a central issue in the calculation of the performances of solid propellant rocket engines. Since the beginning of the development of these engines, a variety of methods have been used to address this problem. The first methods used were purely analytical and could only be applied to simple geometries, although the skill of some researchers led them to solve complex cases of industrial interest. The use of the first digital calculators, to automate calculation, and numerical methods in modern computers applied to differential equations, which adequately describe the kinematics of the free surface, has put the problem of burnback in a state of remarkable technological maturity. Also, a series of phenomenological methods have recently been developed, which use specific properties of the solution, like the principle of minimum time or Piobert's statement, which obtain interesting results but are difficult to generalize to problems with non-uniform recession velocity.

The most general and fruitful methods lie in solving the Eikonal equation which, as shown in this paper, is obtained from the detailed analysis of the process. Although the direct resolution of the equation was addressed early, at the beginning of this century, giving rise to powerful and versatile methods, during the last twenty years the developments have led to solving the burnback problem using the so-called Level Set Method. LSM-based calculations solve a Hamilton-Jacobi equation, using a signed level function, to get the solution robustly and reliably, without limitations in the geometries to analyze nor in the recession velocity distributions. However, this strategy is oversized for the burnback problem. LSM is a procedure that solves much more general problems than burnback but enjoys great popularity because it is used in a very wide range of free-boundary problems and with applications in very different fields. From a broad efficiency point of view, the burnback problem must be solved using the Eikonal equation on an unstructured discretization of the propellant volume, so that it is possible to address any geometric complication that the design problem of a solid-propellant rocket engine requires. The method is computationally efficient, especially when compared with other kinds of analyses that need to be addressed in the design of a solid-propellant rocket engine (e.g. structural or internal aerodynamic calculations). The reason is that only one unknown needs to be solved and the meshing does not need the sophistication of a CFD mesh.

This paper develops the basic theory of propagation of the combustion front, carries out a critical review of the existing literature on burnback analysis, highlights the ability of analytical methods solving very general problems of, for example, bipropellants, and shows the power and versatility of the integration of the Eikonal equation, using simple time marching for the solution of any grain design problem.

# 8. Acknowledgements

This study has been carried out as part of the PILUM project (Proyecto de Investigación de tecnologías para Lanzador, Ubicado en plataforma aérea, de Micro y nano satélites) promoted by INTA (Instituto Nacional de Tecnología Aeroespacial Esteban Terradas), an autonomous agency of the Spanish public administration responsible for the aerospace and defense technologies research and especially from the support received from Tcol. Jesús Sánchez, head of the Department of Rockets and Orthotronics at INTA-Marañosa Campus. It has also received partial support from the Scholarship-Collaboration program of the Spanish Ministry of Education and Science and in part from a similar program sponsored by Universidad Politécnica de Madrid.

# 9. References

[1] D. E. Coats, G. R. Nickerson, A. L. Dang, and S. S. Dunn, "Solid performance program (SPP)," in *23rd AIAA/SAE/ASME/ASEE Joint Propulsion Conference, San Diego, CA, AIAA Paper 87-1701*, 1987.

[2] F. R. Billheimer, J. S.; Wagner, "The Morphological Continuum in Solid Propellant Grain Design," in *Propulsion Re-Entry Physics*, Elsevier, 1970, pp. 152–187.

[3] G. Thibodaux Jr, J. G., Swain, R. L., Wright, *Analytical and experimental studies of spherical solid-propellant rocket motors.* Washington: NACA RM L57G12a, 1957.

[4] M. W. Stone, "Slotted Tube Grain Design," *ARS J.*, vol. 31, no. 2, pp. 223–228, 1961, doi: 10.2514/8.5435.

[5] C. Tola and M. Nikbay, "Internal ballistic modeling of a solid rocket motor by analytical burnback analysis," *J. Spacecr. Rockets*, vol. 56, no. 2, pp. 498–516, 2019, doi: 10.2514/1.A34065.

[6] A. Rafique, Amer F.; Zeeshan, Qasim; Kamran and L. Guozhu, "A new paradigm for star grain design and optimization," *Aircr. Eng. Aerosp. Technol.*, vol. 87, no. 5, pp. 476–482, 2015, doi: 10.1108/AEAT-07-2013-0141.

[7]   A. Kamran, L. Guozhu, J. Godil, Z. Siddique, Q. Zeeshan, and A. F. Rafique, "Design and performance optimization of Finocyl Grain," *AIAA Model. Simul. Technol. Conf.*, no. August, pp. 1–10, 2009, doi: 10.2514/6.2009-6234.

[8]   N. Y. Shapiro, Ya. M.; Mazing, G. Yu.; Prudnikov, "THEORY OF SOLIO FUEL ROCKET ENGINES," 1969.

[9]   S. Krishnan and T. K. Bose, "Design of Multi-Propellant Star Grains for Solid Propellant Rockets.," *Def. Sci. J.*, vol. 30, no. 1, pp. 21–30, 1980, doi: 10.14429/dsj.30.6407.

[10]  S. S. Dunn and D. E. Coats, "3-D grain design and ballistic analysis using the SPP97 code," *33rd Jt. Propuls. Conf. Exhib.*, pp. 1–14, 1997, doi: 10.2514/6.1997-3340.

[11]  D. E. Coats, J. C. French, S. S. Dunn, and D. R. Berker, "Improvements to the Solid Performance Program (SPP)," in *39th AIAA/ASME/SAE/ASEE Joint Propulsion Conference and Exhibit, Huntsville, AL, AIAA Paper 2003-4504*, 2003, no. July, doi: 10.2514/6.2003-4504.

[12]  D. E. Coats and A. L. Dang, "Improvements to the solid performance program (SPP'12) and a review of nozzle performance predictions," *50th AIAA/ASME/SAE/ASEE Jt. Propuls. Conf. Cleveland, OH, AIAA Pap. 2014-3804*, pp. 1–8, 2014, doi: 10.2514/6.2014-3804.

[13]  S. Scippa, "Propellant Grain Design," 1988.

[14]  G. Püskülcü and A. Ulas, "3-D grain burnback analysis of solid propellant rocket motors: Part 2 - modeling and simulations," *Aerosp. Sci. Technol.*, vol. 12, no. 8, 2008, doi: 10.1016/j.ast.2008.01.002.

[15]  A. Mahjub, Q. Azam, M. Z. Abdullah, and N. M. Mazlan, "Cad-based 3d grain burnback analysis for solid rocket motors," 2020, doi: 10.1007/978-981-15-4756-0_28.

[16]  A. Abdelaziz and L. Guozhu, "Two dimensional star grain optimization method using genetic algorithm," 2018, doi: 10.1109/IBCAST.2018.8312216.

[17]  K. O. Reddy and K. M. Pandey, "Burnback Analysis of 3-D Star Grain Solid Propellant," *Int. J. Adv. Trends Comput. Sci. Eng.*, vol. 2, no. 1, pp. 215–223, 2013.

[18]  A. Kamran and L. Guozhu, "Design and optimization of 3D radial slot grain configuration," *Chinese J. Aeronaut.*, vol. 23, no. 4, pp. 409–414, 2010, doi: 10.1016/S1000-9361(09)60235-1.

[19]  A. Abdelaziz and L. Guozhu, "Three Dimensional Modified Star Grain Design and Burnback Analysis," *Int. J. Model. Optim.*, vol. 7, no. 3, 2017.

[20]  P. Noh, W. F.; Woodward, "SLIC (Simple Line Interface Calculation)," 1976.

[21]  N. Mashayek, F;Farzad, H.; Ashgriz, "A Geometry Independent Technique for Solid Propellant Grain Design," *Proc. Inst. Mech. Eng. Part G J. Aerosp. Eng.*, vol. 210, pp. 209–220, 1996, doi: 10.1243/PIME_PROC_1996_210_365_02.

[22]  R. J. Hejl and S. D. Heistert, "Solid Rocket Motor Grain Burnback Analysis Using Adaptive Grids," *J. Propuls. Power*, vol. 11, no. 5, pp. 1006–1011, 1995.

[23]  W. Ki, T. Ko, S. Kim, and W. Yoon, "3D grain burnback analysis using the partial interface tracking method," *Aerosp. Sci. Technol.*, vol. 68, 2017, doi: 10.1016/j.ast.2017.04.023.

[24]  M. A. Willcox, M. Q. Brewster, K. C. Tang, and D. S. Stewart, "Solid propellant grain design and burnback simulation using a minimum distance function," *J. Propuls. Power*, vol. 23, no. 2, pp. 465–475, 2007, doi: 10.2514/1.22937.

[25]  M. A. Willcox, M. Q. Brewster, K. C. Tang, D. S. Stewart, and I. Kuznetsov, "Solid rocket motor internal ballistics simulation using three-dimensional grain burnback," *J. Propuls. Power*,

vol. 23, no. 3, pp. 575–584, 2007, doi: 10.2514/1.22971.

[26] P. REN *et al.*, "Solid rocket motor propellant grain burnback simulation based on fast minimum distance function calculation and improved marching tetrahedron method," *Chinese J. Aeronaut.*, vol. 34, no. 4, pp. 208–224, Apr. 2021, doi: 10.1016/j.cja.2020.08.052.

[27] A. Javed, I. A. Sundaram, and D. Chakraborty, "Internal ballistic code for solid rocket motors using minimum distance function for grain burnback," *Def. Sci. J.*, vol. 65, no. 3, 2015, doi: 10.14429/dsj.65.8304.

[28] Y. Liu, J. Sui, Y. Zhao, F. Bao, and W. Hui, "Large Scale Parallel Algorithms for 3D Grain Burnback Analysis of Solid Propellant Rocket Motors," in *Proceedings of the 22nd International Conference on Industrial Engineering and Engineering Management 2015*, 2016.

[29] Y. Ata, D. F. Kurtulus, and O. U. Arkun, "Development of a 3D Grain Burnback Simulation Tool for Solid Rocket Motors," in *Advances in Sustainable Aviation*, T. H. Karakoç, C. O. Colpan, and Y. \cSöhret, Eds. Cham: Springer International Publishing, 2018, pp. 65–90.

[30] Y. H. Hwang and C. H. Chiang, "Simple surface-tracking methods for grain burnback analysis," *J. Guid. Control. Dyn.*, vol. 38, no. 6, 2015, doi: 10.2514/1.B35682.

[31] J. A. Osher, S.;Sethian, "Fronts propagating with curvature-dependent speed: Algorithms based on Hamilton-Jacobi formulations," *J. Comput. Phys.*, vol. 79, no. 1, pp. 12–49, 1988, doi: 10.1016/0021-9991(88)90002-2.

[32] S. Osher, R. Fedkiw, and K. Piechor, *Level Set Methods and Dynamic Implicit Surfaces*, vol. 57, no. 3. 2004.

[33] J. A. Sethian, *Level Set Meyhods and Fast marching Metethods*. Cambridge University Press, 1996.

[34] C. Yildirim and H. Aksel, "Numerical Simulation of the Grain Burnback in Solid Propellant Rocket Motor," 2005, no. July, doi: 10.2514/6.2005-4160.

[35] F. Qin, H. Guoqiang, L. Peijin, and L. Jiang, "Algorithm study on burning surface calculation of solid rocket motor with complicated grain based on level set methods," *Collect. Tech. Pap. - AIAA/ASME/SAE/ASEE 42nd Jt. Propuls. Conf.*, vol. 6, no. July, pp. 4476–4484, 2006, doi: 10.2514/6.2006-4774.

[36] E. Cavallini, "Modeling and Numerical Simulation of Solid Rocket Motors Internal Ballistics," Sapienza University of Rome, 2008.

[37] Y. Liu, K. Yin, F. Bao, Y. Liu, and E. Wu, "Efficient simulation of grain burning surface regression," *Appl. Mech. Mater.*, vol. 466–467, no. 1, pp. 314–318, 2012, doi: 10.4028/www.scientific.net/AMR.466-467.314.

[38] A. P. Lorente, "Development of the Quasi-3D model for the grain burnback analysis of SRM's," in *Proceedings of the International Astronautical Congress, IAC*, 2013, vol. 9.

[39] W. Sullwald, F. Smit, A. Steenkamp, and W. Rousseau, "Solid rocket motor grain burn back analysis using level set methods and monte-carlo volume integration," *49th AIAA/ASME/SAE/ASEE Jt. Propuls. Conf.*, vol. 1 PartF, 2013, doi: 10.2514/6.2013-4087.

[40] C. W. Rousseau, S. F. Steyn, W. Sullwald, E. R. De Kock, G. J. F. Smit, and J. H. Knoetze, "Rapid solid rocket motor design," *49th AIAA/ASME/SAE/ASEE Jt. Propuls. Conf.*, vol. 1 PartF, pp. 1–12, 2013, doi: 10.2514/6.2013-3789.

[41] D. H. Wang, Y. Fei, F. Hu, and W. H. Zhang, "An integrated framework for solid rocket motor grain design optimization," *Proc. Inst. Mech. Eng. Part G J. Aerosp. Eng.*, vol. 228, no. 7, pp.

1156–1170, 2014, doi: 10.1177/0954410013486589.

[42] G. L. Mejia, R. J. Rocha, L. Rocco, S. R. Gomes, K. Iha, and J. A. F. F. Rocco, "Solid rocket motor burn simulation considering complex 3D propellant grain geometries," *52nd AIAA/SAE/ASEE Jt. Propuls. Conf. 2016*, pp. 1–6, 2016, doi: 10.2514/6.2016-5098.

[43] M. H. Tshokotsha, "Internal Ballistic Modelling of Solid Rocket Motors Using Level Set Methods for Simulating Grain Burnback by," 2016.

[44] R. Wei, F. Bao, Y. Liu, and W. Hui, "Combined Acceleration Methods for Solid Rocket Motor Grain Burnback Simulation Based on the Level Set Method," *Int. J. Aerosp. Eng.*, vol. 2018, no. May, 2018, doi: 10.1155/2018/4827810.

[45] S. Mesgari, M. Bazazzadeh, and A. Mostofizadeh, "Finocyl grain design using the genetic algorithm in combination with adaptive basis function construction," *Int. J. Aerosp. Eng.*, vol. 2019, 2019, doi: 10.1155/2019/3060173.

[46] S. H. Oh, H. J. Lee, and T. S. Roh, "Development of a hybrid method in a 3-D numerical burnback analysis for solid propellant grains," *Aerosp. Sci. Technol.*, vol. 106, p. 106103, 2020, doi: 10.1016/j.ast.2020.106103.

[47] A. Chiapolino, F. Fraysse, and R. Saurel, "A Method to Solve Hamilton–Jacobi Type Equation on Unstructured Meshes," *J. Sci. Comput.*, vol. 88, no. 1, pp. 1–43, 2021, doi: 10.1007/s10915-021-01517-9.

[48] E. Saintout, A. Le Roux, D. Ribereau, and P. Perrin, "ELEA - A tool for 3D surface regression analysis in propellant grains," in *25th Joint Propulsion Conference AIAA/ASME/SAE/ASEE, Monterey, CA, July 10-12, AIAA 89-2782*, 1989, p. ., doi: 10.2514/6.1989-2782.

[49] D. Le Roux, A. Y.; NaMah, G. S.;Riberau, "Numerical Model for Propellant Grain Rurning Surface Recesion," in *Mathematical Modeling in Combustion and Related Topics*, 1988, pp. 505–514, doi: 10.1007/978-94-009-2770-4_35.

[50] G. Uhrig, B. Ducourneau, and P. Liesa, "Computer aided preliminary design of propellant grains for solid rocket motors," *AIAA/ASME/SAE/ASEE 23rd Jt. Propuls. Conf. 1987*, 1987, doi: 10.2514/6.1987-1734.

[51] F. Dauch and D. Ribéreau, "A software for SRM grain design and internal ballistics evaluation: PIBAL®," *38th AIAA/ASME/SAE/ASEE Jt. Propuls. Conf. Exhib.*, vol. 2002–4299, 2002.

[52] P. Le Breton, D. Ribéreau, F. Godfrey, R. Abgrall, and S. Augoula, "SRM performance analysis by coupling bidimensional surface burnback and pressure field computations," 1998, doi: 10.2514/6.1998-3968.

[53] D. Ribéreau, P. Le Breton, and E. Giraud, "SRM 3D surface burnback computation using mixes stratification deduced from 3D grain filling simulation," in *35th Joint Propulsion Conference and Exhibit*, 1999, no. June, doi: 10.2514/6.1999-2802.

[54] J. A. Sethian, *Fast Marching Methods and Level Set Methods*. 1998.

[55] N. Ashgriz and J. Y. Poo, "FLAIR: Flux line-segment model for advection and interface reconstruction," *J. Comput. Phys.*, vol. 93, no. 2, pp. 449–468, 1991, doi: 10.1016/0021-9991(91)90194-P.

[56] R. Bertacin, F. Ponti, and A. Annovazzi, "A new three-dimensional ballistic model for Solid Rocket Motor non-homogeneous combustion," *48th AIAA/ASME/SAE/ASEE Jt. Propuls. Conf. Exhib. 2012*, no. August, pp. 1–13, 2012, doi: 10.2514/6.2012-3974.

[57] F. Ponti, S. Mini, L. Fadigati, V. Ravaglioli, A. Annovazzi, and V. Garreffa, "Effects of

inclusions on the performance of a solid rocket motor," *Acta Astronaut.*, vol. 189, no. May, pp. 283–297, 2021, doi: 10.1016/j.actaastro.2021.08.030.

[58] C. Yildirim, "Analysis Of Grain Burnback And Internal Flow In Solid Propellant Rocket Motor In 3-dimensions." METU.

[59] R. L. Nowack, "Wavefronts and solutions of the eikonal equation," *Geophys. J. Int.*, vol. 110, no. 1, pp. 55–62, 1992, doi: 10.1111/j.1365-246X.1992.tb00712.x.

[60] D. Gueyffier *et al.*, "Accurate computation of grain burning coupled with flow simulation in rocket chamber," *J. Propuls. Power*, vol. 31, no. 6, pp. 1761–1776, 2015, doi: 10.2514/1.B35736.

[61] K. A. Toker, "Three-Dimensional Retarding Walls and Flow in their Vicinity," 2004.

[62] M. Barrere, A. Jaumotte, B. Fraeijs de Vaubeke, and J. Vandenkerckhove, *Rocket propulsion*. 1960.

[63] NASA, "Solid Propellant Grain Gesign and Internal Ballistics," no. March. 1972.

[64] M. G. Crandall, H. Ishii, and P. L. Lions, "User's guide to viscosity solutions of second order partial differential equations," *Bull. Am. Math. Soc.*, vol. 27, no. 1, pp. 1–67, 1992, doi: 10.1090/S0273-0979-1992-00266-5.

[65] M. G. Crandall and P.-L. Lions, "Viscosity Solutions of Hamilton-Jacobi Equations," *Trans. Am. Math. Soc.*, vol. 277, no. 1, pp. 1–42, 1983.

[66] M. G. Crandall and P. L. Lions, "Two Approximations of Solutions of Hamilton-Jacobi Equations," *Math. Comput.*, vol. 43, no. 167, p. 1, 1984, doi: 10.2307/2007396.

[67] S. Osher and J. A. Sethian, "Fronts Propagating with Curvature Dependent Speed: Algorithms Based on Hamilton-Jacobi," *J. Comput. Phys.*, vol. 79, no. 1, pp. 12–49, 1988.

[68] A. Harten, B. Engquist, S. Osher, and S. R. Chakravarthy, "Uniformly high order accurate essentially non-oscillatory schemes, III," *J. Comput. Phys.*, vol. 71, no. 2, pp. 231–303, 1987, doi: 10.1016/0021-9991(87)90031-3.

[69] C.-W. Shu and S. Osher, "Efficient lmplementation of Essentially Non-oscillatory Shock-Capturing Schemes," *J. Comput. Phys.*, vol. 77, pp. 439–471, 1988.

[70] G. S. Jiang and D. Peng, "Weighted ENO schemes for Hamilton-Jacobi equations," *SIAM J. Sci. Comput.*, vol. 21, no. 6, pp. 2126–2143, 2000, doi: 10.1137/S106482759732455X.

[71] S. Osher and C.-W. Shu, "High-Order Essentially Nonoscillatory Schemes for Hamilton-Jacobi Equations Author," *SIAM J. Numer. Anal.*, vol. 28, no. 4, pp. 907–922, 1991.

[72] R. Abgrall, "Numerical Discretization of the First-Order Hamilton- Jacobi Equation on Triangular Meshes," *Commun. Pure Appl. Math.*, vol. XLIX, pp. 1339–1373, 1996.

[73] M. Bardi and L. C. Evans, "On Hopf's formulas for solutions of Hamilton-Jacobi equations," *Nonlinear Anal.*, vol. 8, no. 11, pp. 1373–1381, 1984, doi: 10.1016/0362-546X(84)90020-8.

[74] S. Augoula and R. Abgrall, "High order numerical discretization for Hamilton-Jacobi equations on triangular meshes," *J. Sci. Comput.*, vol. 15, no. 2, pp. 197–229, 2000, doi: 10.1023/A:1007633810484.

[75] Y.-T. Zhang and C.-W. Shu, "High-order schemes for Hamilton-Jacobi equations on triangular meshes," *SIAM J. Sci. Comput.*, vol. 24, no. 3, pp. 1005–1030, 2003, doi: 10.1016/j.cam.2003.09.051.

[76] X. G. Li, W. Yan, and C. K. Chan, "Numerical schemes for Hamilton-Jacobi equations on

unstructured meshes," *Numer. Math.*, vol. 94, no. 2, pp. 315–331, 2003, doi: 10.1007/s00211-002-0418-9.

[77] R. Abgrall and J. D. Benamou, "Big ray-tracing and eikonal solver on unstructured grids: Application to the computation of a multivalued traveltime field in the Marmousi model," *Geophysics*, vol. 64, no. 1, pp. 230–239, 1999, doi: 10.1190/1.1444519.

[78] J. A. Sethian and A. Vladimirsky, "Fast methods for the Eikonal and related Hamilton-Jacobi equations on unstructured meshes," *Proc. Natl. Acad. Sci. U. S. A.*, vol. 97, no. 11, pp. 5699–5703, 2000, doi: 10.1073/pnas.090060097.

[79] L. Huang, C. Shu, and M. Zhang, "Numerical boundary conditions for the fast sweeping high order WENO methods for solving the Eikonal equation," *J. Comput. Math.*, vol. 26, no. 3, pp. 336–346, 2008.